\documentclass[twocolumn,twocolappendix]{aastex63}

\newcommand*\patchAmsMathEnvironmentForLineno[1]{
  \expandafter\let\csname old#1\expandafter\endcsname\csname #1\endcsname
  \expandafter\let\csname oldend#1\expandafter\endcsname\csname end#1\endcsname
  \renewenvironment{#1}
  {\linenomath\csname old#1\endcsname}
  {\csname oldend#1\endcsname\endlinenomath}}
  \newcommand*\patchBothAmsMathEnvironmentsForLineno[1]{
  \patchAmsMathEnvironmentForLineno{#1}
  \patchAmsMathEnvironmentForLineno{#1*}}
  \AtBeginDocument{
  \patchBothAmsMathEnvironmentsForLineno{equation}
  \patchBothAmsMathEnvironmentsForLineno{align}
  \patchBothAmsMathEnvironmentsForLineno{flalign}
  \patchBothAmsMathEnvironmentsForLineno{alignat}
  \patchBothAmsMathEnvironmentsForLineno{gather}
  \patchBothAmsMathEnvironmentsForLineno{multline}
}

\usepackage{lineno}

\usepackage[normalem]{ulem}

\usepackage{amsmath}

\received{2021 September 29}
\revised{2021 December 6}
\accepted{2022 January 4}
\shorttitle{Measuring off-axis BBH spins with Plus-era detectors}
\shortauthors{Knee et al.}

\begin{document}

\title{Prospects for measuring off-axis spins of binary black holes with Plus-era gravitational-wave detectors}

\correspondingauthor{Alan M. Knee}
\email{aknee@phas.ubc.ca}

\author[0000-0003-0703-947X]{Alan M. Knee}
\affiliation{Department of Physics \& Astronomy, University of British Columbia, Vancouver, BC V6T 1Z1, Canada}

\author[0000-0003-0316-1355]{Jess McIver}
\affiliation{Department of Physics \& Astronomy, University of British Columbia, Vancouver, BC V6T 1Z1, Canada}


\author[0000-0003-4059-4512]{Miriam Cabero}
\affiliation{Department of Physics \& Astronomy, University of British Columbia, Vancouver, BC V6T 1Z1, Canada}



\begin{abstract}
The mass and spin properties of binary black holes (BBHs) inferred from their gravitational-wave signatures reveal important clues about how these systems form. BBHs originating from isolated binary evolution are expected to have spins preferentially aligned with their orbital angular momentum, whereas there is no such preference in binaries formed via dynamical assembly. The fidelity with which near-future gravitational-wave detectors can measure off-axis spins will have implications for the study of BBH formation channels. In this work, we examine the degree to which the Advanced LIGO Plus (A+) and Advanced Virgo Plus (AdV+) interferometric detectors can measure both aligned and misaligned spins. We compare spin resolution between the LIGO-Virgo network operating at either A+/AdV+ (``Plus'') sensitivity or Advanced-era design (``Design'') sensitivity using simulated BBH gravitational-wave signals injected into synthetic detector noise. The signals are distributed over the mass-spin parameter space of likely BBH systems, accounting for the effects of precession and higher-order modes. We find that the Plus upgrades yield significant improvements in spin estimation for systems with unequal masses and moderate or large spins. Using simulated signals modelled after different types of hierarchical BBH mergers, we also conclude that the Plus detector network will yield substantially improved spin estimates for 1G+2G binaries compared to the Design network.\\
\end{abstract}



\section{Introduction}\label{sec:intro}


Since the first direct detection of a binary black hole (BBH), GW150914 \citep{PhysRevLett.116.061102}, gravitational waves from coalescing compact binaries have revealed an emerging population of heavy stellar-mass BBHs. To date, the LIGO Scientific and Virgo Collaborations (LVC) have reported a total of 90 gravitational-wave candidates\footnote{This tally reflects the number of events determined to have at least a 50\% probability of being astrophysical.} from the first three observing runs \citep{LIGOScientific:2018mvr, Abbott:2020niy, LIGOScientific:2021usb, LIGOScientific:2021qlt, LIGOScientific:2021djp} of the Advanced LIGO \citep{TheLIGOScientific:2014jea} and Advanced Virgo \citep{TheVirgo:2014hva} interferometers. The majority of detected events are most likely associated with BBHs, with total masses ranging from $14\,M_\odot$ to $184\,M_\odot$. The sensitivities of gravitational-wave detectors have steadily improved with each observing run, and the upcoming fourth (O4) and fifth (O5) observing runs planned for this decade are expected to add hundreds of new detections \citep{Aasi:2013wya, Baibhav:2019gxm}.

The mass and spin properties of BBH sources inferred from their gravitational-wave emissions test our theories of how these systems are formed. For instance, the unusually massive component black holes (BHs) associated with GW190521 \citep{Abbott:2020tfl} potentially lie in the putative upper mass gap\footnote{GW190521 may also have an asymmetric mass ratio, which would change the component mass estimates \citep{Nitz:2020mga, Capano:2021etf, Estelles:2021jnz}.}, approximately between $65$--$140\,M_\odot$. Late-stage stellar evolution theory predicts that stars cannot produce remnant BHs in this mass range due to the onset of pair-instability supernovae (PSN) \citep{Belczynski:2016jno, Woosley_2017}, implying that GW190521 might be of dynamical origin \citep{Kimball:2020qyd, Romero-Shaw:2020thy, Bustillo:2020syj, Gayathri:2020coq}. This hypothesis is also supported by weak evidence of orthogonal (in-plane) spin components relative to the orbit. Although it is currently uncertain whether GW190521 points to a novel population of BBHs \citep{Abbott:2020mjq}, it has highlighted how the mass and spin information encoded in gravitational waves can be used to understand how BBHs form.

In general, BBHs can be grouped into one of two proposed categories of formation channels: isolated binary evolution, and dynamical assembly. In the isolated scenario, a BBH is formed from a highly evolved binary star, independent of the gravitational influence of any outside sources. This channel may involve a common envelope phase \citep{Bethe:1998bn, 8147515, 2002ApJ...572..407B, Belczynski:2016obo}, or proceed via chemically homogeneous evolution \citep{Mandel:2015qlu, deMink:2016vkw}. Alternatively, a BBH can be formed through repeated dynamical interactions. Unlike the isolated channel, the dynamical channel is viable only in dense stellar environments where close encounters are frequent, such as in the cores of globular clusters and nuclear star clusters \citep{Gerosa:2021mno, 1993Natur.364..423S, Portegies_Zwart_2002, Rodriguez:2016avt, Rodriguez:2016kxx, Belczynski:2020bnq}. Accretion disks around active galactic nuclei are also expected to drive BBH mergers, as BBHs can have their inspirals accelerated by interacting with the gaseous disk \citep{2012MNRAS.425..460M, 2017ApJ...835..165B}. Hierarchical triples represent another possible channel, in which a binary merger is driven by a perturbing tertiary mass through the Lidov-Kozai mechanism \citep{Lim:2020cvm, 2020ApJ...903...67M}.

Formation channels are distinguished by the imprints they leave on the properties of BBHs, including the masses and spins of their components, as well as orbital eccentricity. BBHs formed via the isolated channel are expected to have spins preferentially aligned with the orbital angular momentum of the system as a result of tidal interactions or mass transfer \citep{Kalogera:1999tq, Bogdanovic:2007hp}. In dynamical binaries, the spins should be oriented isotropically, resulting in misaligned systems \citep{Vitale:2015tea, Talbot:2017yur, Farr:2017uvj, Farr:2017gtv}. Spin misalignment also induces general relativistic spin-precession, causing the orbital plane to precess about the total angular momentum of the system. Precession introduces further periodic structure into the gravitational-wave signal by modulating its amplitude and phase evolution \citep{PhysRevD.49.6274, Kidder:1995zr}. Dynamical binaries may also have eccentric orbits if there is insufficient time for the orbit to circularize via emission of gravitational waves before the merger \citep{Romero-Shaw:2019itr}, providing another method for constraining formation channels \citep{Zevin:2021rtf}. 

The dynamical channel can facilitate hierarchical mergers, in which the remnant BHs of previous mergers proceed to form a new (dynamical) binary and merge once again. This process could potentially stratify the BH population into different ``generations'', resulting in various classes of dynamical binaries with unique mass and spin properties depending on how the generations are paired up \citep{Gerosa:2018wbw, Rodriguez:2017pec, Rodriguez:2019huv, Sedda:2020vwo, Callister:2021fpo}. BHs that originated from collapsed stars are thus termed ``first generation'' (1G) BHs, while the remnant of the merger of two 1G BHs is ``second generation'' (2G), and so on. Hierarchical mergers are expected to have higher average masses and spins, since these higher-generation BHs inherit much of the mass and angular momentum of their progenitor binaries \citep{Fishbach:2017dwv, Gerosa:2017kvu}, and could theoretically populate the upper mass gap \citep{Kimball:2020opk, Gerosa:2021mno}. Precise measurements of both the masses and spins will therefore play a critical role in the study of these formation channels with gravitational waves.


Although our ability to measure the spins of gravitational-wave sources is limited by current detector sensitivities, future technological upgrades to these detectors present an opportunity to achieve higher-quality spin measurements, potentially allowing us to make more definitive statements about the formation histories of detected events. Work is currently underway to bring the Advanced LIGO and Advanced Virgo detectors up to their design sensitivity settings. This iteration of the LIGO-Virgo network is planned to be superseded by Advanced LIGO Plus (A+) and Advanced Virgo Plus (AdV+). The proposed A+/AdV+ upgrades feature a suite of modifications, including improved mirror coatings and the implementation of frequency-dependent light squeezing, which will achieve a broad-spectrum reduction in noise across the LIGO-Virgo observing band \citep{PhysRevLett.123.231107, McCuller:2020yhw}, greatly enhancing the overall sensitivity of the detectors. The A+ network is projected to have a detection range of 2.5 Gpc for a $30\,M_\odot+30\,M_\odot$ BBH; about 1.6 times the range of Advanced LIGO at design sensitivity, and twice that of the last observing run. Similarly, AdV+ will have a BBH range of roughly a factor of two times its design sensitivity (see \citet{Aasi:2013wya} for an outline of future observing prospects).

Motivated by the expected sensitivity improvements provided by these upgrades, we analyze multiple sets of simulated gravitational-wave signals from BBH sources and compare spin resolution between the ``Design'' and ``Plus'' networks, where ``Design'' refers to the three-detector LIGO-Virgo network operating at Advanced-era design sensitivity, and ``Plus'' refers to the same network with A+/AdV+ sensitivity. By comparing spin resolution between these two networks, as well as examining any systematic correlations with their source properties, we evaluate the implications that these near-future detector upgrades have for the study of BBH formation channels. Previous studies have helped us to understand how current and future gravitational-wave detectors can resolve the spins of BBH systems and the signatures of misalignment/precession \citep{Biscoveanu:2021nvg, Vitale:2014mka, Vitale:2016avz, Vitale:2016icu, Bustillo:2016gid, Stevenson:2017dlk, Gerosa:2018wbw, Pratten:2020igi, Green:2020ptm, Kalaghatgi:2020gsq, Kimball:2020opk}. This work expands on this body of literature by explicitly comparing the ability of these two networks to recover the spin information of a wide range of sources, and determining the degree to which our spin measurements improve as a result of the Plus upgrades.

The remainder of this paper is organized as follows: In Section \ref{sec:bg}, we review how the spin properties of BBHs are parameterized in gravitational-wave analyses; in Section \ref{sec:method}, we outline our parameter estimation and injection methods; in Section \ref{sec:results}, we present and discuss the results of our injection analyses, and finally we conclude in Section \ref{sec:disc}. 

\section{Precessing binary black holes}\label{sec:bg}

In general, inspiralling BBHs are fully characterized by eight intrinsic quantities, assuming quasi-circular orbits. These are the primary (heavier) and secondary (lighter) component masses, $m_1\geq m_2$, as measured in the source frame\footnote{The masses measured in the frame of the detector are cosmologically redshifted, and are related to the ``true'' source-frame masses by a redshift factor, $m_{1,2}^{\rm det}=(1+z)m_{1,2}$. In this work, any time a mass is written without the superscript, it is implied to be a source-frame mass.}, plus six parameters for the two three-dimensional spin vectors of the component BHs, $\mathbf{S}_{1,2}$ \citep{Farr:2014qka}. It is common practice to parameterize the binary in terms of: the chirp mass, $\mathcal{M}= (m_1m_2)^{3/5}/(m_1+m_2)^{1/5}$; mass ratio, $q= m_1/m_2\geq 1$; dimensionless spin magnitudes\footnote{Note some authors write the spin magnitudes as $\chi_{1,2}$.}, $a_{1,2}= c|\mathbf{S}_{1,2}|/(G m_{1,2}^2)\leq 1$; tilt angles, $t_{1,2}=\arccos(\hat{\mathbf{L}}\cdot\hat{\mathbf{S}}_{1,2})$, between the spins and the orbital angular momentum, $\mathbf{L}$; and two additional angles, $\phi_{\rm JL}$ and $\phi_{12}$, that define the in-plane orientations of the spins. Since the spin tilts and orientation angles are not conserved as the system evolves, they must be defined at a reference frequency. In this work, we quote the values of these parameters at $f=20$ Hz, near the lower limit of the LIGO-Virgo sensitivity band. Additionally, seven extrinsic parameters are needed to describe the location and orientation of the source relative to the detector coordinate system. These include the luminosity distance to the source, $D_L$, sky location (right ascension, $\alpha$, and declination, $\delta$), polarization angle, $\psi$, and an inclination parameter (measured at a reference frequency), $\theta_{\rm JN}$, between the total system angular momentum, $\mathbf{J}=\mathbf{L}+\mathbf{S}_1+\mathbf{S}_2$, and the detector line-of-sight vector, $\mathbf{N}$. Finally, we also need to specify the time of coalescence (in the geocentric frame), $t_c$, and the orbital phase at the chosen reference frequency, $\phi_{\rm ref}$.

As a consequence of their relatively subtle effects on waveform morphology, the spin magnitudes $a_{1,2}$ are usually difficult to constrain individually, and are sensitive to prior assumptions \citep{Vitale:2017cfs, Biscoveanu:2020are, LIGOScientific:2018mvr}. The leading-order effect of the spins on the gravitational waveform is characterized by the mass-weighted sum of the aligned projections of the spins, known as the effective aligned spin \citep{Damour:2001tu, Racine:2008qv, Ajith:2009bn}:
\begin{equation}\label{eq:chieff}
\chi_{\rm eff} = \frac{qa_1\cos t_1+a_2\cos t_2}{1+q}\,,\quad -1\leq\chi_{\rm eff}\leq 1\,.
\end{equation}
Due to its effect on the inspiral phase \citep{Ajith:2009bn, Baird:2012cu}, $\chi_{\rm eff}$ is typically the most well-measured combination of the spins \citep{Vitale:2016avz}. Qualitatively, the spins can either accelerate or delay the onset of the merger \citep{Campanelli:2006uy, Healy:2018swt, Ng:2018neg}. Aligned-spin sources ($\chi_{\rm eff}>0$) complete a greater number of orbits prior to merging than non-spinning sources due to the need to dissipate more angular momentum, thereby extending the inspiral. Anti-aligned-spin sources ($\chi_{\rm eff}<0$) instead complete fewer orbits, and thus have a shorter observable inspiral.

Other spin effects arise when the spins are misaligned with the orbit, which causes the system to precess. The orbital angular momentum will rotate about the total angular momentum, which remains roughly fixed under the assumption that the orbit precesses on a timescale significantly shorter than that of the inspiral \citep{Kidder:1995zr}. The approximate effect of precession on the waveform is captured by the effective precession spin, defined as \citep{Schmidt:2014iyl}:
\begin{equation}\label{eq:chip}
\chi_{\rm p} = {\rm max}\bigg(a_1\sin t_1, \frac{4+3q}{q(4q+3)}a_2\sin t_2\bigg)\,,\quad 0\leq\chi_{\rm p}\leq 1\,.
\end{equation}
It is clear from this formula that $\chi_{\rm p}$ will be non-zero if either of the spins are misaligned ($\sin t_{1,2}\neq 0$). In practice, this parameter does not capture all of the precession behaviour. Alternative methods of describing precession are given in e.g.~\citet{Thomas:2020uqj, Fairhurst:2019srr, Fairhurst:2019vut, Gerosa:2020aiw, Akcay:2020qrj}. Precession introduces periodic modulations of the phase and amplitude of the emitted gravitational-wave signal \citep{PhysRevD.49.6274, Kidder:1995zr}, as it causes the orientation of the orbit to change relative to the line-of-sight. This breaks the equatorial symmetry of the system, modifying the mode structure via mode-mixing \citep{Schmidt:2010it, Khan:2019kot} and breaking parameter degeneracies \citep{Lang:1900bz}, including a well-known mass-spin degeneracy that afflicts non-precessing systems \citep{Baird:2012cu, Chatziioannou:2014coa, Tiwari:2018qch, Pratten:2020igi}. 


\section{Method}\label{sec:method}

In order to systematically compare spin resolution between the Design and Plus detector networks, we generated a multitude of simulated gravitational-wave signals (``injections'') from BBH sources, and attempted to recover their parameters using these two networks. To create the simulated signals, we used the \textsc{IMRPhenomXPHM} waveform approximant \citep{Pratten:2020ceb, Garcia-Quiros:2020qpx}; a fully-precessing, quasi-circular inspiral-merger-ringdown model of coalescing BBHs that supports higher-order multipoles beyond the dominant $(\ell,|m|)=(2,2)$ quadrupole modes. The same approximant is used for recovering the parameters later. The signals were injected into realizations of stationary Gaussian noise\footnote{This is a reasonably close approximation to real LIGO-Virgo noise.} coloured by the noise power spectral density (PSD) of the detector. The analyzed strain data is given by $d(t)=n(t)+h(t)$, where $n(t)$ and $h(t)$ correspond to the noise and signal strain, respectively \citep{LIGOScientific:2019hgc}. We assumed a three-detector network consisting of the LIGO Hanford, LIGO Livingston, and Virgo interferometers, operating at their Design and Plus settings.

\subsection{Parameter estimation}

We performed parameter estimation using the formalism of Bayesian statistical inference \citep{Christensen:2001cr, 2019PASA...36...10T}. With this approach, constraints on waveform parameters are derived from their posterior probability distributions (``posteriors'') conditioned on the data. Given a waveform model $h$ dependent on source parameters $\boldsymbol{\theta}=(\mathcal{M},q,\ldots)$, and interferometric strain data $\mathbf{d}$, the posteriors are inferred using Bayes' theorem,
\begin{equation}\label{eq:bayes}
P(\boldsymbol{\theta}|\mathbf{d},h) = \frac{L(\mathbf{d}|\boldsymbol{\theta},h)\pi(\boldsymbol{\theta}|h)}{Z(\mathbf{d}|h)}\,.
\end{equation}
Here, the likelihood function, $L(\mathbf{d}|\boldsymbol{\theta},h)$, is the probability of observing the data conditioned on the parameter values, while $\pi(\boldsymbol{\theta}|h)$ is the prior distribution, representing our initial assumptions about the parameter distributions. The denominator is a normalizing factor called the evidence, and is defined as
\begin{equation}\label{eq:zevid}
Z(\mathbf{d}|h) = \int L(\mathbf{d}|\boldsymbol{\theta},h)\pi(\boldsymbol{\theta}|h)\,d\boldsymbol{\theta}\,.
\end{equation}
For the likelihood, we employ a stationary Gaussian noise model weighted by the detector PSD, expressed in the frequency domain as \citep{Veitch:2008ur, Veitch:2014wba}
\begin{equation}\label{eq:like}
\ln L(\mathbf{d}|\boldsymbol{\theta},h)\propto -\sum_k \frac{2|\tilde{d}_k-\tilde{h}_k(\boldsymbol{\theta})|^2}{S_n(f_k)T}\,,
\end{equation}
where $f_k$ are the frequency bins, $\tilde{d}_k$ and $\tilde{h}_k$ are respectively the discrete Fourier transforms of the strain data and waveform model, $S_n(f)$ is the noise PSD, and $T$ is the duration of the analyzed data segment. Once the posterior distribution is known, constraints on individual or pairs of parameters can be obtained by marginalizing the posterior over the other remaining parameters. For a network of multiple detectors, the full log-likelihood involves another sum over each detector.

\begin{figure}[t!]
\epsscale{1.15}
\plotone{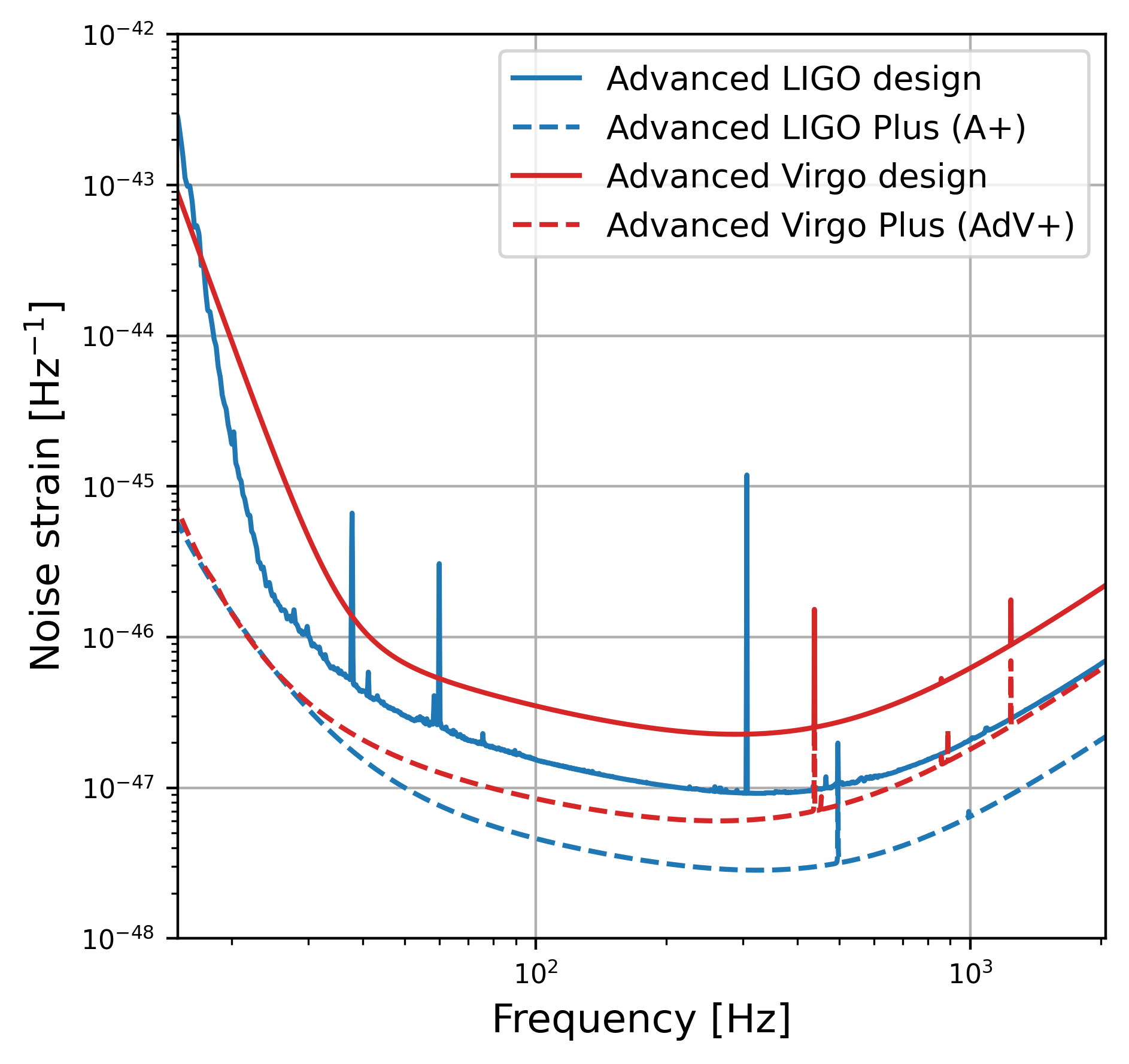}
\caption{Noise curves for the Advanced LIGO and Virgo detectors at their Design and Plus sensitivity settings. The vertical axis is the PSD of the noise in units of Hz$^{-1}$. A downward shift in the noise curve implies higher sensitivity. The solid (dashed) curves make up what we call the Design (Plus) network. \label{fig:psds}}
\end{figure}

The creation of the injections and subsequent analyses were carried out using the publicly-available \textsc{Bilby} pipeline \citep{Ashton:2018jfp, Romero-Shaw:2020owr}, which supports a variety of Monte Carlo and nested sampling methods to explore the parameter space. In particular, we infer the posteriors using a  \textsc{Bilby}-specific implementation of the nested sampling algorithm \textsc{Dynesty} \citep{2020MNRAS.493.3132S, 2004AIPC..735..395S, 10.1214/06-BA127}. To improve convergence times, we use a distance-marginalized likelihood using the techniques of \citet{Singer:2015ema}. Each injection is analyzed assuming Design sensitivity and then again with Plus so that we have two sets of posteriors to compare. The PSDs for the Design and Plus networks are obtained from \citet{OReilly2018}, and are shown in Figure \ref{fig:psds}.

\subsection{Injections}\label{sec:grids}

The injection sets are divided into two main parts. In the first part, we create $10\times 10$ grids of injections spaced discretely in two parameters, which we list here.
\begin{itemize}
	\item \emph{Mass-spin}: This grid surveys a broad region of the mass and spin parameter space, providing a general picture of how the Design and Plus networks compare in terms of resolving BBHs with in-plane spins, and how this depends on the mass ratios and spins. Injections are spaced discretely in mass ratio, $q\in [1, 10]$, and spin magnitude, $a_{1,2}\in [0, 0.9]$, with $a_1=a_2$. The spin tilts, $t_{1,2}$, are both fixed to $90^\circ$, and there is a fixed total mass of $M=65\,M_\odot$ (similar to GW150914).
	\item \emph{Spin tilt 1G+2G}: This grid examines the full range of spin tilt angles, with fixed mass and spins chosen to roughly emulate the properties of a merger between 1G and 2G BHs. The 2G primary component is assumed to be twice the mass of the 1G secondary. Additionally, the primary component has high spin inherited from the angular momentum of the progenitor binary, while the secondary component is a stellar remnant with low spin. Injections are again spaced discretely, but in the two cosine spin tilts  $\cos t_{1,2}\in [-1, 1]$, at fixed mass ratio, $q=2$, and total mass, $M=100\,M_\odot$. The spin magnitudes are fixed at $a_1=0.7$ and $a_2=0.1$, the former of which is a reasonable estimate based on the predicted spin distribution of hierarchically formed 2G BHs \citep{Fishbach:2017dwv, Gerosa:2017kvu}. We do not experiment with different total masses, as preliminary testing showed little systematic correlation between total mass and spin recovery for $M>80\,M_\odot$.
	\item \emph{Spin tilt 2G+2G}: This grid is similar to the 1G+2G grid, but simulating mergers between two 2G BHs. The two components have comparable masses, and both have large spins by virtue of being the products of previous mergers. Injections are spaced discretely in the two cosine spin tilts, $\cos t_{1,2}\in [-1, 1]$, at fixed mass ratio, $q=1.1$, and total mass, $M=100\,M_\odot$. The spin magnitudes are fixed at $a_1=0.7$ and $a_2=0.8$.
\end{itemize}
All extrinsic and remaining intrinsic parameters are randomly sampled from appropriate distributions: the two intrinsic spin angles, $\phi_{\rm JL}$ and $\phi_{12}$, are drawn from a uniform distribution between $0$ and $2\pi$; the polarization angle, $\psi$, is uniform between $0$ and $\pi$; the orbital reference phase, $\phi_{\rm ref}$, is uniform between $0$ and $2\pi$; the inclination angle, $\theta_{\rm JN}$, is isotropic on the sphere; the sky location, $(\alpha,\delta)$, is isotropic on the sky; and the luminosity distance is drawn from a distribution such that the redshift is uniform in comoving volume,
\begin{equation}\label{eq:zprior}
P(z) \propto \frac{1}{1+z}\frac{dV_c}{dz}\,,
\end{equation}
where the extra redshift factor is to account for the transformation between source and detector-frame time. Distances are drawn out to $\sim 2$ Gpc so that the bulk of the simulated events have Design network SNRs in the range of $8$--$60$. The Plus network SNRs are $1.8$--$2.1$ times larger. 

The analyzed bandwidth is truncated at a minimum frequency of 15 Hz, which we deemed appropriate for Design/Plus sensitivity. We also ensure each injection is at least ``detectable'' by applying an optimal signal-to-noise ratio (SNR) \citep{DelPozzo:2014cla} cutoff to reject quiet signals. The cutoff imposes that each injection has a network SNR\footnote{The network SNR is the SNR observed in each detector added in quadrature.} across the three detectors of at least 8, and an SNR of at least 5 in two of the detectors at Design sensitivity. 

For the analyses of the two spin tilt grids, uniform priors are adopted for the detector-frame chirp mass, $\mathcal{M}^{\rm det} \in [30, 70] \,M_\odot$, and for the mass-spin grid we use $\mathcal{M}^{\rm det} \in [5, 50]\,M_\odot$. For all three grids we use the following priors: inverted mass ratio, $q^{-1}\in [0.05, 1]$; spin magnitudes, $a_{1,2}\in [0, 0.95]$; and coalescence time, $t_c\in [-0.1,0.1]$ (where each injection has $t_c=0$). Isotropic priors are used for the remaining angular parameters, as shown in Table I of \citet{Ashton:2018jfp}. 

In the second part of our study, we analyze an injection set consisting of 200 signals with masses and spins sampled from the astrophysical population models described in \citet{Abbott:2020gyp}. The primary masses are drawn from the ``power law + peak'' model; a compound distribution consisting of a descending power law with a smooth taper and a Gaussian peak:
\begin{align}
P(m_1) &= [(1-\lambda)\xi(m_1|-\alpha, m_{\rm max}) \nonumber \\
&+\lambda G(m_1|\mu_m,\sigma_m^2)]S(m_1|m_{\rm min},\delta_m)\,,
\end{align}
The secondary masses are obtained assuming a power law for the mass ratio, $P(m_2)\propto (m_2/m_1)^{\beta_q} S(m_2|m_{\rm min},\delta_m)$. Since we also want to examine how the detector networks resolve the spins of both aligned and misaligned systems, the spin properties are sampled from the ``default spin'' model. Here, the two spin magnitudes are drawn from identical beta distributions, and the spin tilts are drawn from a mixture of two populations:
\begin{equation}
P(\cos t_1,\cos t_2) = \frac{1-\zeta}{4}+ \zeta G_t(\cos t_1,\cos t_2|\sigma_t^2)\,.
\end{equation}
The first term corresponds to an isotropic tilt distribution that models dynamical binaries, where any combination of spin angles is equally likely. In isolated binaries, the spins are favoured to be in a nearly aligned configuration, which is modelled by a truncated Gaussian with mean $(\cos t_1,\cos t_2)=(1,1)$ and variance $\sigma_t^2$ for both spin tilts. For the definitions and astrophysical motivations behind these models, see e.g.~\citet{Kovetz:2016kpi, Talbot:2017yur, Wysocki:2018mpo, Talbot:2018cva, Abbott:2020gyp}. 

We use the median posterior values for the parameters reported in \citet{Abbott:2020gyp}. The extrinsic and remaining intrinsic parameters are sampled in the same manner as the injection grids described earlier, and we apply the same SNR cuts to ensure that all the signals qualify as detectable with Design sensitivity. The minimum frequency is still $15$ Hz, but we extend the chirp mass prior to $\mathcal{M}^{\rm det} \in [3, 100]\,M_\odot$ to accommodate the wider range of masses. Events are simulated out to $\sim 1$ Gpc which achieves roughly the same SNR distribution as the injection grids\footnote{The Plus network will be able to observe far more distant events, even beyond $z=1$, but for purposes of determining how well we can measure off-axis spins it is not necessary to simulate farther signals.}.

\subsection{Constraint statistics}\label{sec:cost}

The information contained in a posterior distribution can be condensed down to a variety of metrics, depending on what aspects of the posteriors are considered most interesting. One commonly used metric is the Bayesian credible interval (CI), which is defined as the interval $I$ such that some parameter $\theta$ falls in that interval with probability $\mathcal{C}$,
\begin{equation}
\int_I P(\theta|\mathbf{d},h)\,d\theta = \mathcal{C}\,.
\end{equation}
Note that the posterior here has been marginalized over the other parameters. One can define the CI as an interval about the median. In what follows, we quote the 90\% ($\mathcal{C}=0.9$) CIs, so the intervals are bounded between the 5th and 95th percentiles of the posteriors. To quantify the precision of the CI, we can use their widths. We denote these widths (at 90\% credibility) by $\Sigma_{90}(\theta)$. For two posteriors obtained at Design and Plus sensitivity, we can compare them using the relative change in $\Sigma_{90}(\theta)$, given by $\Delta_{90}(\theta)=[\Sigma^{\rm Plus}_{90}(\theta)-\Sigma^{\rm Design}_{90}(\theta)]/\Sigma^{\rm Design}_{90}(\theta)$.

The CI only tells us about the variance of the posterior, and contains no information about whether the analysis is recovering the correct parameter values. Given that we work with injections, where the true values of the parameters are known, we would like to incorporate this knowledge when describing the constraints. Furthermore, we inject into Gaussian noise which can cause posteriors to be offset from the simulated values, especially for low-SNR signals near the threshold, so we have reason to expect that the analysis will not always recover the signal parameters accurately. We thus introduce a generalized variance quantity, referred to as a ``cost function'', which averages the squared difference from the injected value (rather than the mean) over the posterior distribution:
\begin{equation}
C(\theta) = \bigg[\int_{-\infty}^\infty P(\theta|\mathbf{d},h)(\theta-\theta_0)^2\,d\theta\bigg]^{1/2}\,,
\end{equation}
where $\theta_0$ is the injected value. If the posterior mean is equal to the injected value, the cost function is equal to the posterior standard deviation. Should the posterior be offset from the injected value, it will be penalized with a higher cost. Again, we can compare the cost at the two sensitivities by looking at the relative change in the cost function, $\Delta C(\theta) = [C^{\rm Plus}(\theta)-C^{\rm Design}(\theta)]/C^{\rm Design}(\theta)$. Further discussion about the cost function is given in Appendix \ref{app:cost}. To facilitate a more faithful comparison between the two sensitivities, the noise realizations are generated with a seed that is unique to each injected signal. Thus, the noise realizations across the two networks retain the same likelihood, but differ through a frequency-dependent amplitude scaling.\footnote{Specifically, this rescaling depends on the relative difference in the noise PSDs between the Design and Plus networks.} 

In the following analyses, we use both the 90\% CIs and $C(\theta)$ to characterize parameter recovery. Most of our results are discussed in terms of the latter, but we also include the posterior widths for selected results in Appendix \ref{app:widths}.

\section{Results}\label{sec:results}

In this section, we present the results of our parameter estimation analyses. We focus first on the mass-spin and two spin tilt injection grids, noting any trends between the injected source properties and differences in spin recovery between the Design and Plus networks. We then turn our attention towards the injections sampled from population models. 

The precision of parameter estimation depends largely upon the SNR of the signal. The SNR is determined by the signal amplitude (a function of the mass, distance, and inclination of the system) and detector sensitivity. Intrinsic properties, namely the mass ratio and the spins, also play a significant role. For instance, the effective aligned spin, $\chi_{\rm eff}$, will modify the duration of the inspiral phase, potentially yielding better spin estimates. Spin estimation also benefits from higher mass ratios, as found in e.g.~\citet{Vitale:2014mka, Vitale:2016avz}. Systems with misaligned spins will also precess, de-correlating the mass ratio and effective spin and leading to improved constraints on both the masses and spins. Meanwhile, the relative improvement between the Design and Plus networks will depend on differences in the PSDs, as well as the frequency content of the signals.

\subsection{Mass-spin grid}

\begin{figure}[t!]
\epsscale{1.2}
\plotone{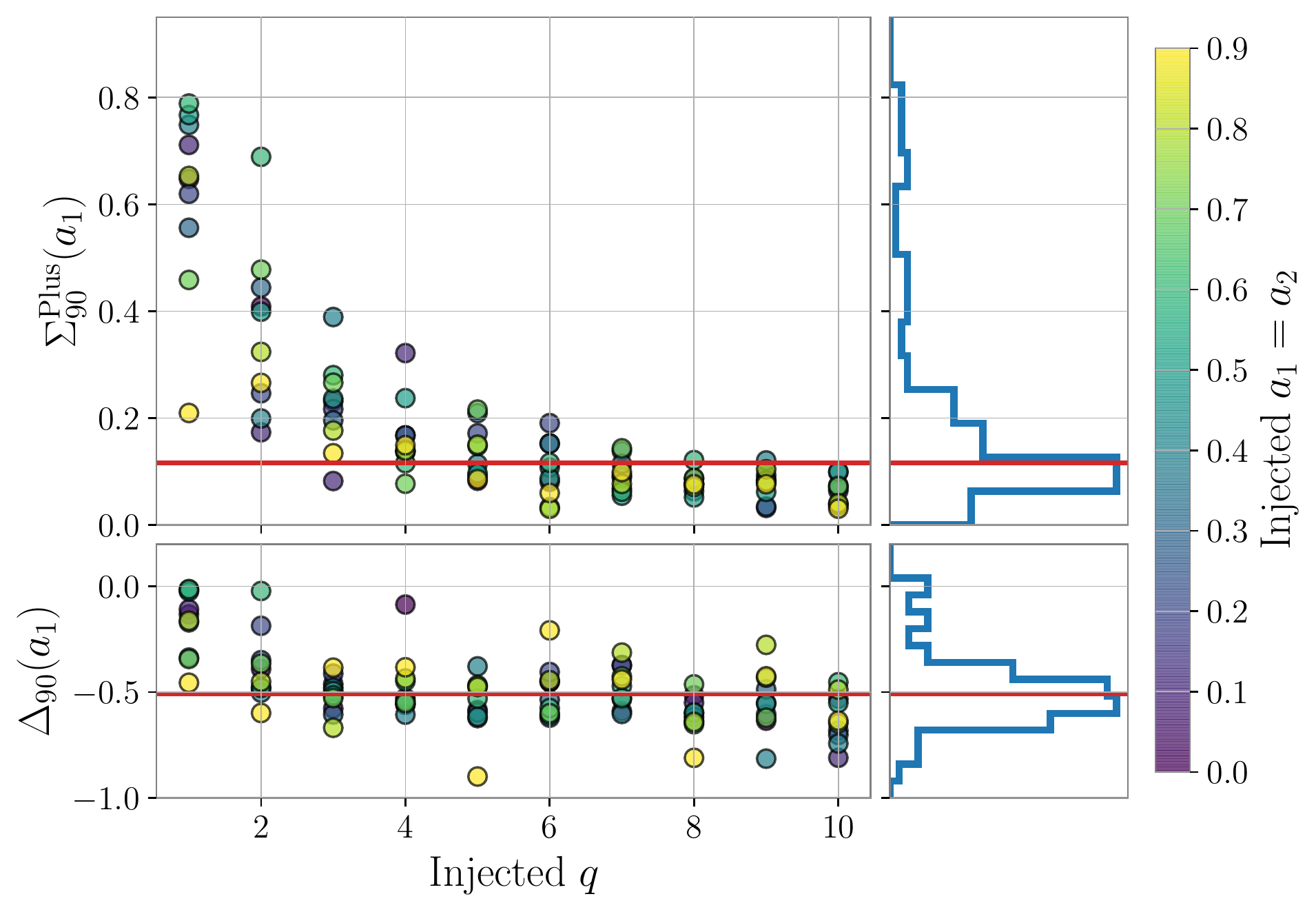}
\caption{Distribution of constraints on the primary spin magnitude, $a_1$, as a function of the mass ratio, $q$, for the mass-spin injection grid. {\it Top:} Widths of the 90\% CIs at Plus sensitivity, $\Sigma_{90}^{\rm Plus}(a_1)$. The histogram shows how the widths are distributed. The colourbars indicate the injected value of $a_1=a_2$. The red line marks the median value of $\Sigma_{90}^{\rm Plus}(a_1)$. {\it Bottom:} Relative change in CI widths between Design and Plus sensitivity, $\Delta_{90}(a_1)$, also with a histogram. The median $\Delta_{90}(a_1)$ is marked by a red line. \label{fig:mass_spin_a1}}
\end{figure}

\begin{figure}[t!]
\epsscale{1.2}
\plotone{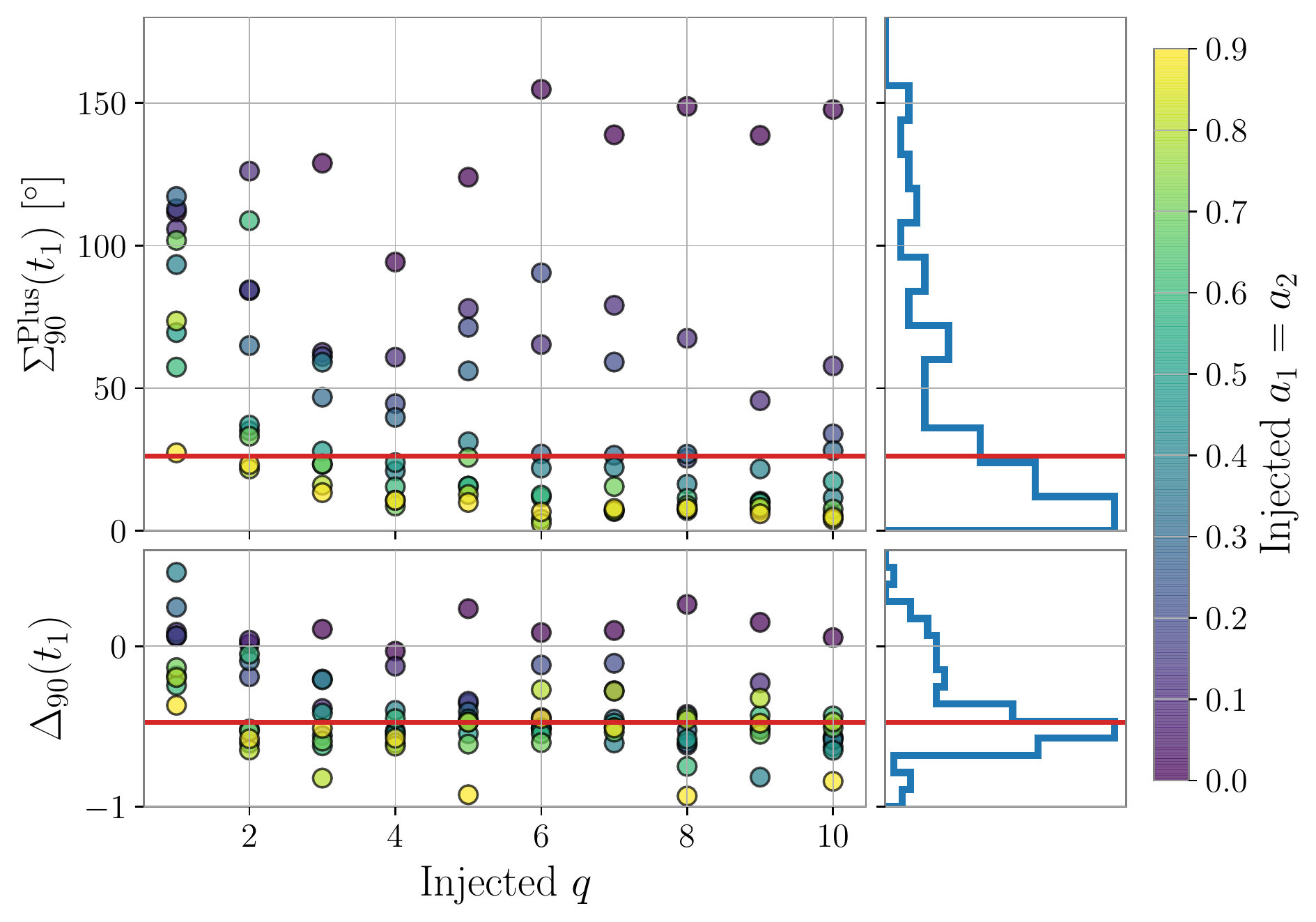}
\caption{Same as Figure \ref{fig:mass_spin_a1}, now showing CIs for the primary tilt, $t_1$. \label{fig:mass_spin_t1}}
\end{figure}

Starting with the mass-spin grid, we show the constraints on the primary spin magnitude, $a_1$, and primary tilt, $t_1$, in Figures \ref{fig:mass_spin_a1}--\ref{fig:mass_spin_t1} as functions of the mass ratio. These figures include the widths of the 90\% CIs, $\Sigma_{90}$, for Plus sensitivity and the relative change $\Delta_{90}$ in these widths compared to Design sensitivity. We can see that the uncertainties in the primary spin decrease as the mass ratio increases. The largest uncertainties are found in the equal-mass cases, while for $q\geq 2$ nearly all the injections have $\Sigma_{90}^{\rm Plus}(a_1)<0.5$ and $\Sigma_{90}^{\rm Plus}(t_1)<90^\circ$. This is consistent with how higher mass ratios improve spin estimation as described above. This effect is significantly enhanced when higher-order modes are present, which are amplified by mass asymmetry \citep{LIGOScientific:2020stg}, breaking a degeneracy between the mass ratio and spins. The tilt of the primary spin is also better measured when the spins are larger, as the high-spin injections typically have the smallest $t_1$ uncertainties at fixed $q$. The largest uncertainties for $t_1$ similarly occur in equal-mass systems. Note that even at high mass ratios, we still encounter a few cases with very large $t_1$ uncertainties ($\Sigma_{90}^{\rm Plus}(t_1)\sim 150^\circ$). These injections have $a_1=a_2=0$, where the value of the tilt becomes arbitrary and is thus unconstrained. Otherwise, the uncertainty in $t_1$ for injections with non-zero spins decreases with higher $q$.

\begin{figure}[t!]
\epsscale{1.15}
\plotone{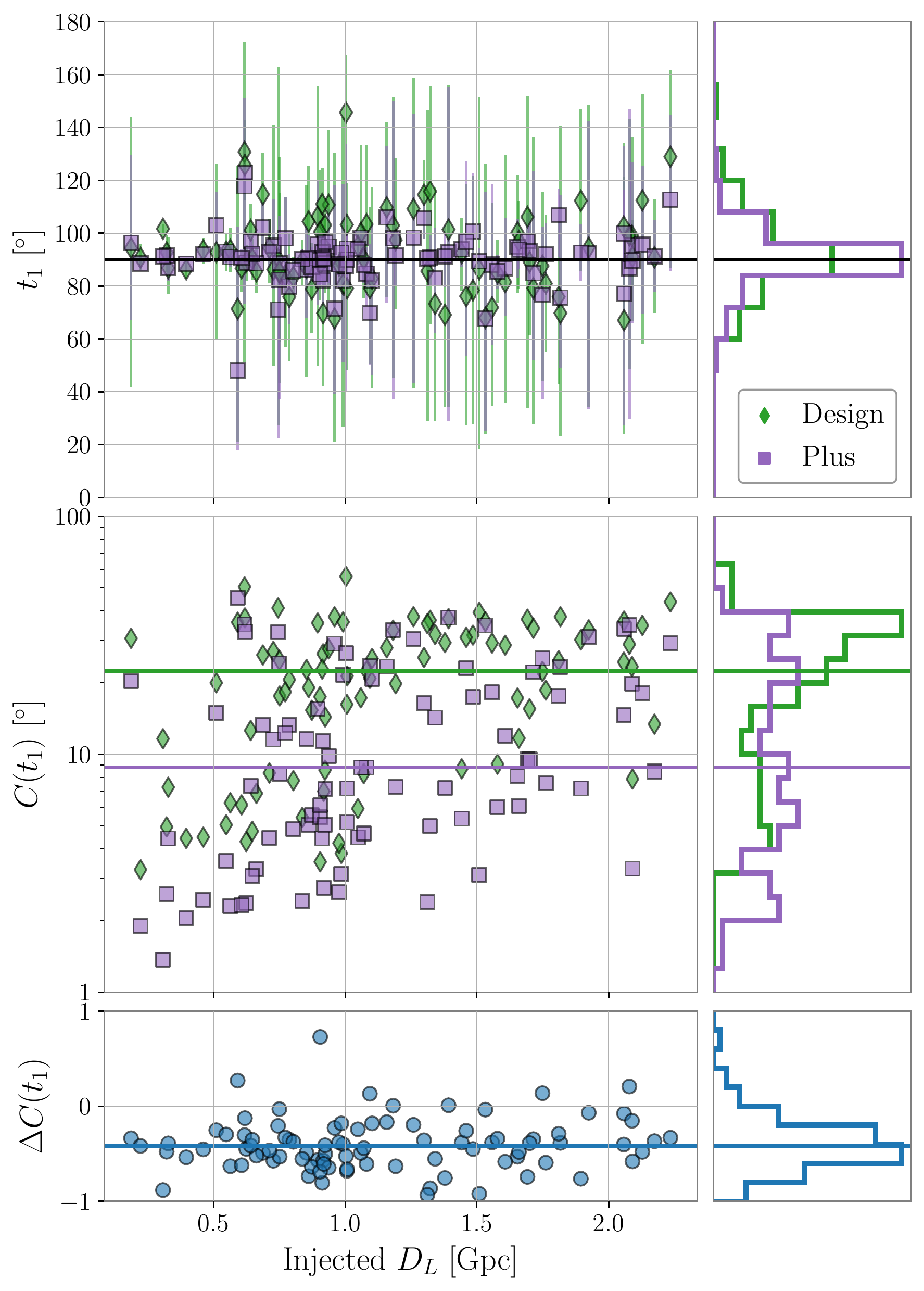}
\caption{Recovery of the primary tilt, $t_1$, as a function of luminosity distance, $D_L$, for the mass-spin injection grid. The ten non-spinning injections are not shown because their tilts are not well-defined. {\it Top:} Posterior medians (squares/diamonds) with errorbars indicating the extent of the 90\% CIs (vertical lines). The histogram shows the distributions of the posterior medians for the Design and Plus networks. The injected value $t_1=90^\circ$ is indicated by the solid horizontal line. {\it Middle:} Cost functions, $C(t_1)$, calculated for each $t_1$ posterior, with histograms showing the distributions for the two networks. The solid horizontal lines are the medians of the distributions. {\it Bottom:} Relative change in cost, $\Delta C(t_1)$, between Design and Plus sensitivity for each injected signal. A horizontal line marks the median. \label{fig:mass_spin_cost_t_1}}
\end{figure}
 
The medians of $\Delta_{90}(a_1)$ and $\Delta_{90}(t_1)$ indicate that the constraints on the primary spin are generally around 50\% narrower at Plus sensitivity compared to Design. The bottom panels of Figures \ref{fig:mass_spin_a1}--\ref{fig:mass_spin_t1} show that the CI widths change less for equal mass systems compared to unequal masses. We also find that $t_1$ is generally better measured with the Plus network for higher spin systems. Of the injections with spins $a_{1,2}\geq 0.3$, the median value of $\Delta_{90}(t_1)$ is $-0.5$, compared to injections with spins of 0.1 or 0.2, where the median is $-0.3$. The constraints between systems with mass ratios $2\leq q\leq 10$ and spins $0.3\leq a_{1,2}\leq 0.9$ are otherwise similar, in that $\Sigma_{90}$ and $\Delta_{90}$ are roughly stable for injections within these ranges. Thus, we conclude that the Plus network will yield the largest improvements in spin estimation for systems with unequal masses and moderate-to-high spins, reducing the uncertainty in our measurements by on average a factor of two.

We do not notice similar behaviour in the secondary spin, as both its magnitude and tilt are typically unconstrained. This is largely because the spin of the lighter component has a smaller effect on the waveform, and is thus harder to measure. We find that constraints on $a_2$ and $t_2$ do not improve with higher mass ratio and spins for the Plus network, and in the few cases where the secondary spin is better measured, it is for the sources with the smallest distances/highest SNRs.

The 90\% CIs for the primary tilt $t_1$ against the luminosity distance to the source are shown in Figure \ref{fig:mass_spin_cost_t_1}. The injections with zero spins ($a_1=a_2=0$) are excluded from this figure. We find that the constraints at both detector sensitivities are generally consistent with the injected value, with $t_1=90^\circ$ contained in the 90\% CI in roughly 90\% of cases. The middle panel of Figure \ref{fig:mass_spin_cost_t_1} shows the cost function $C(t_1)$ for the primary tilt posteriors at Design and Plus sensitivity. The cost function reveals that the medians for the Plus network are usually closer to the injected value. We can see in the lower panel of Figure \ref{fig:mass_spin_cost_t_1} that $\Delta C(t_1)<0$ for most (91\%) of the non-spinning injections, confirming that we generally obtain more accurate estimates with the Plus network compared to Design.


\subsection{Spin tilt 1G+2G grid}

\begin{figure}[t!]
\epsscale{1.15}
\plotone{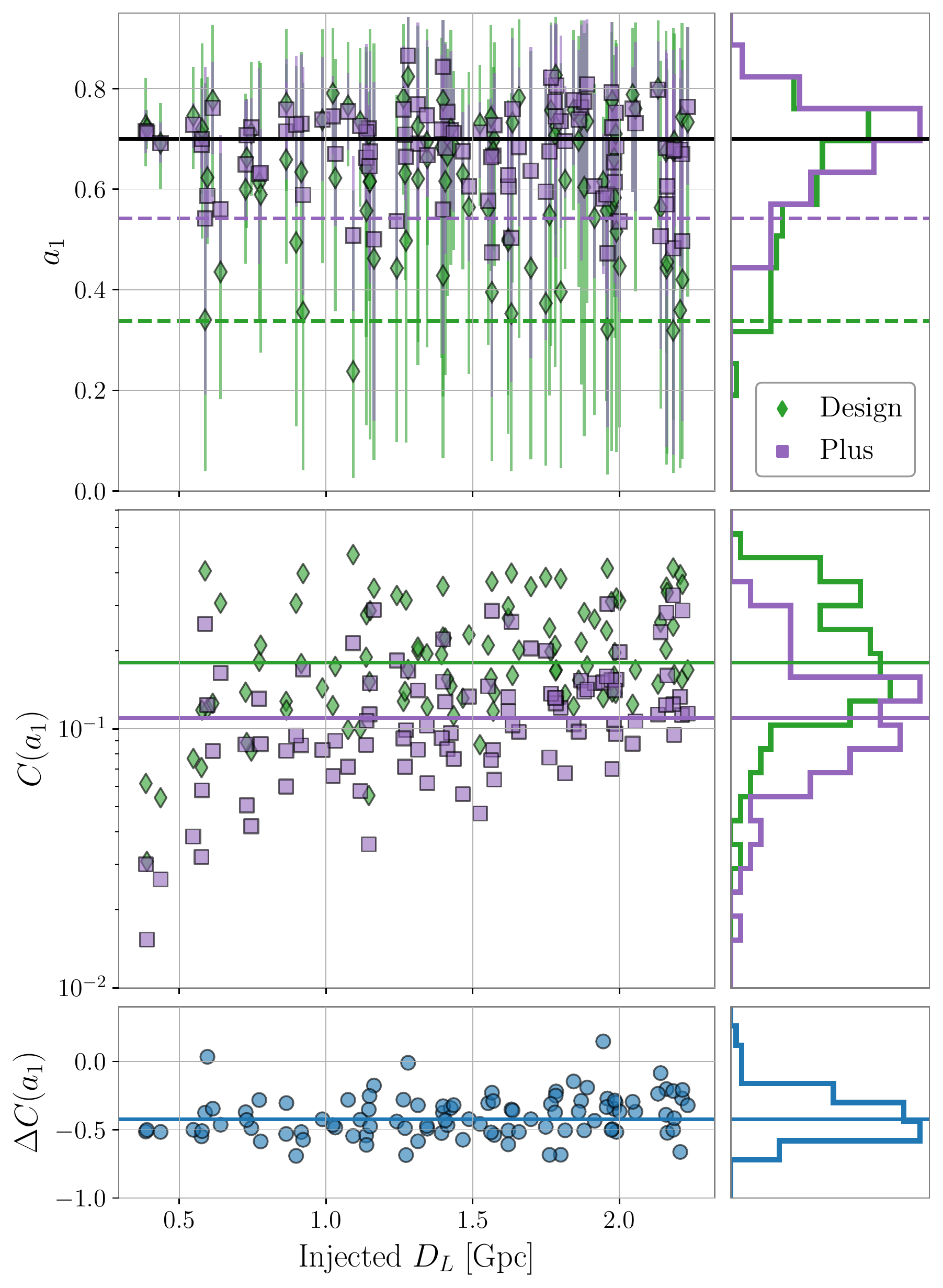}
\caption{Recovery of the primary spin magnitude, $a_1$, as a function of luminosity distance, $D_L$, for the spin tilt 1G+2G injection grid. {\it Top:} Posterior medians (squares/diamonds) with errorbars indicating the extent of the 90\% CIs (vertical lines). The histogram shows the distributions of the posterior medians for the Design and Plus networks. The injected value $a_1=0.7$ is indicated by the solid horizontal line. and the dashed horizontal lines show the median 5th percentiles of the CIs. {\it Middle:} Cost functions, $C(a_1)$, calculated for each $a_1$ posterior, with histograms showing the distributions for the two networks. The solid horizontal lines are the medians of the distributions. {\it Bottom:} Relative change in cost, $\Delta C(a_1)$, between Design and Plus sensitivity for each injected signal. A horizontal line marks the median.  \label{fig:1G2G_a1}}
\end{figure}

We find that the Plus network can recover high $a_1$ at approximately three times the rate for Design for BBH systems consistent with a 1G+2G hierarchical merger. The constraints obtained for $a_1$ are shown in Figure \ref{fig:1G2G_a1}. In the top panel, we show the posterior medians and 90\% CIs, along with the distributions of the medians for the two networks. Although both networks have distributions that peak near the injected value of $a_1=0.7$, Design suffers from a longer tail due to a higher occurrence of constraints that underestimate $a_1$. For the Plus network, we can note qualitatively that the constraints are more heavily clustered around the true injected value than Design.

In order to compare the amount of support for a high primary spin, we have drawn the median position of the 5th posterior percentiles (horizontal dashed lines) in the top panel of Figure \ref{fig:1G2G_a1}. At Plus sensitivity this median sits at 0.54, i.e~half of the injections have $a_1$ constrained above 0.54 with at least 95\% credibility, indicating strong support for a high primary spin. In the context of formation channels, spin magnitudes $a_1\gtrsim 0.5$ are astrophysically significant, as they coincide with the predicted spin distribution of hierarchical BHs \citep{Fishbach:2017dwv}. The Design network yields comparable constraints for only 16\% of the injections.

The middle panel of Figure \ref{fig:1G2G_a1} shows the cost function of the $a_1$ posteriors for the 1G+2G grid. The downward shift in the cost distribution of the Plus network relative to Design reinforces that the constraints are generally narrower at Plus sensitivity for these systems, and that the posteriors peaks are indeed closer to the injected value. The lower panel in Figure \ref{fig:1G2G_a1} shows the relative change in cost $\Delta C(a_1)$ across the two sensitivities, and we find the cost function is of order 40\% lower at Plus sensitivity than Design. 

\begin{figure}[t!]
\epsscale{1.15}
\plotone{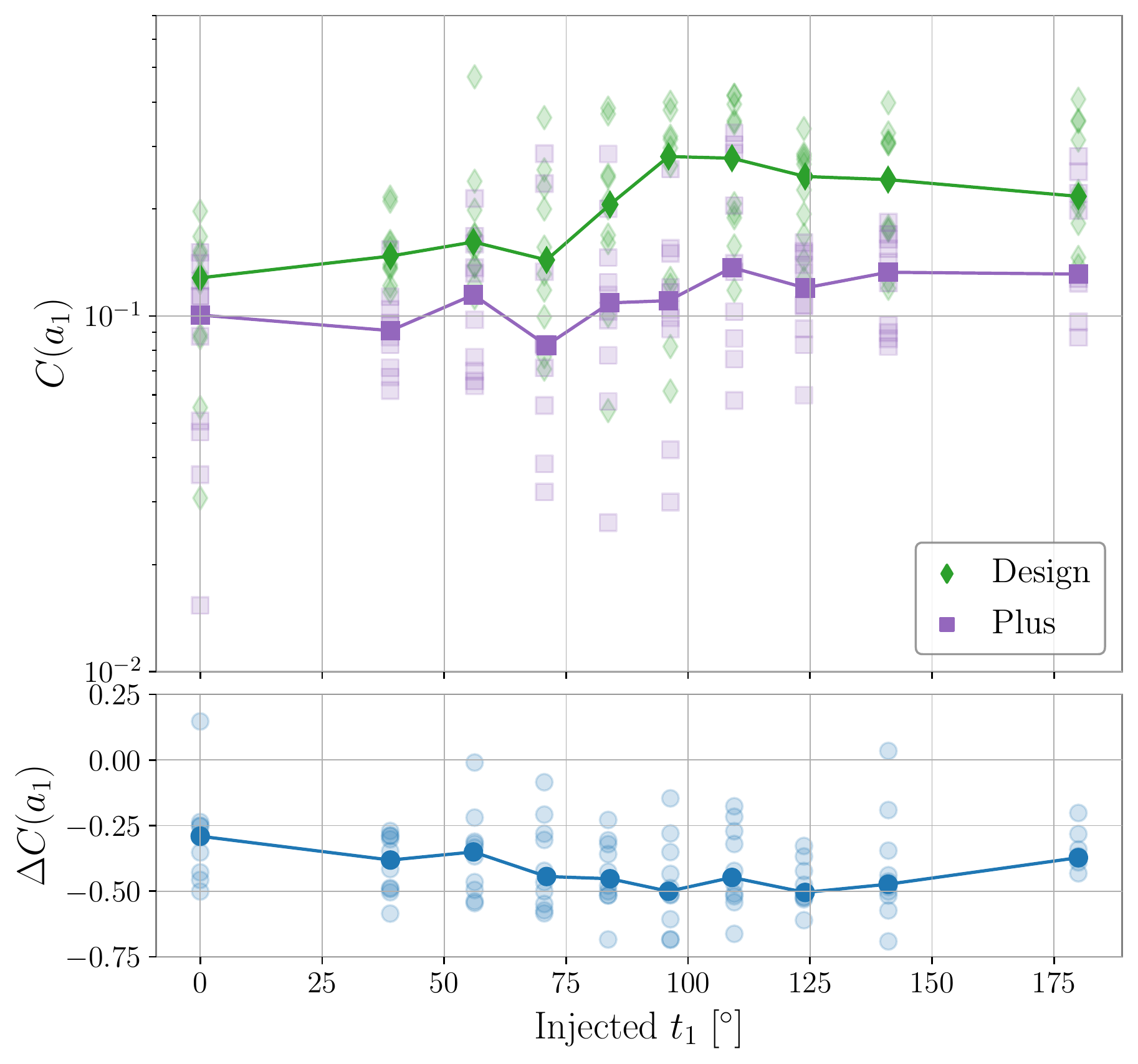}
\caption{Recovery of the primary spin magnitude, $a_1$, as a function of the primary tilt, $t_1$, for the spin tilt 1G+2G injection grid. Transparent points correspond to individual posteriors. For each set of injections with the same value of $t_1$, we indicate the median of the quantity on the vertical axis with an opaque point. {\it Top:} Cost functions, $C(a_1)$, for $a_1$. {\it Bottom:} Relative change in cost, $\Delta C(a_1)$, between Design and Plus sensitivity for each injected signal. \label{fig:1G2G_a1_t1}}
\end{figure}

Section \ref{sec:bg} explained how the inspiral duration is affected by having (anti-)aligned spins. Since much spin/precession information is derived from the inspiral, we expect the spin constraints to be sensitive to the spin tilts. We illustrate this in Figure \ref{fig:1G2G_a1_t1}, where we show the $a_1$ cost function $C(a_1)$ against the injected values of the tilt $t_1$, and highlight the median values of $C(a_1)$ across injections with the same $t_1$. In the bottom panel of Figure \ref{fig:1G2G_a1_t1}, we see that there is greater improvement in the constraints for systems with a primary spin in the range $90^\circ< t_1<150^\circ$, with $|\Delta C(a_1)|$ being twice as large for these cases compared to aligned spins.

1G+2G mergers are characterized by having two components that formed via different channels. Support for large $a_1$ is consistent with the interpretation of the primary component as a 2G BH, and the case for a hierarchical merger could be strengthened if this were combined with convincing evidence of a small secondary spin. However, the secondary spin magnitudes $a_2$ are usually unconstrained at both sensitivities, owing to its small magnitude and the asymmetric mass configuration. At Plus sensitivity, the median 95th percentile for $a_2$ is 0.78. Such a large upper limit suggests we often cannot tell whether the secondary spin is small or not. 

\begin{figure}[t!]
\epsscale{1.15}
\plotone{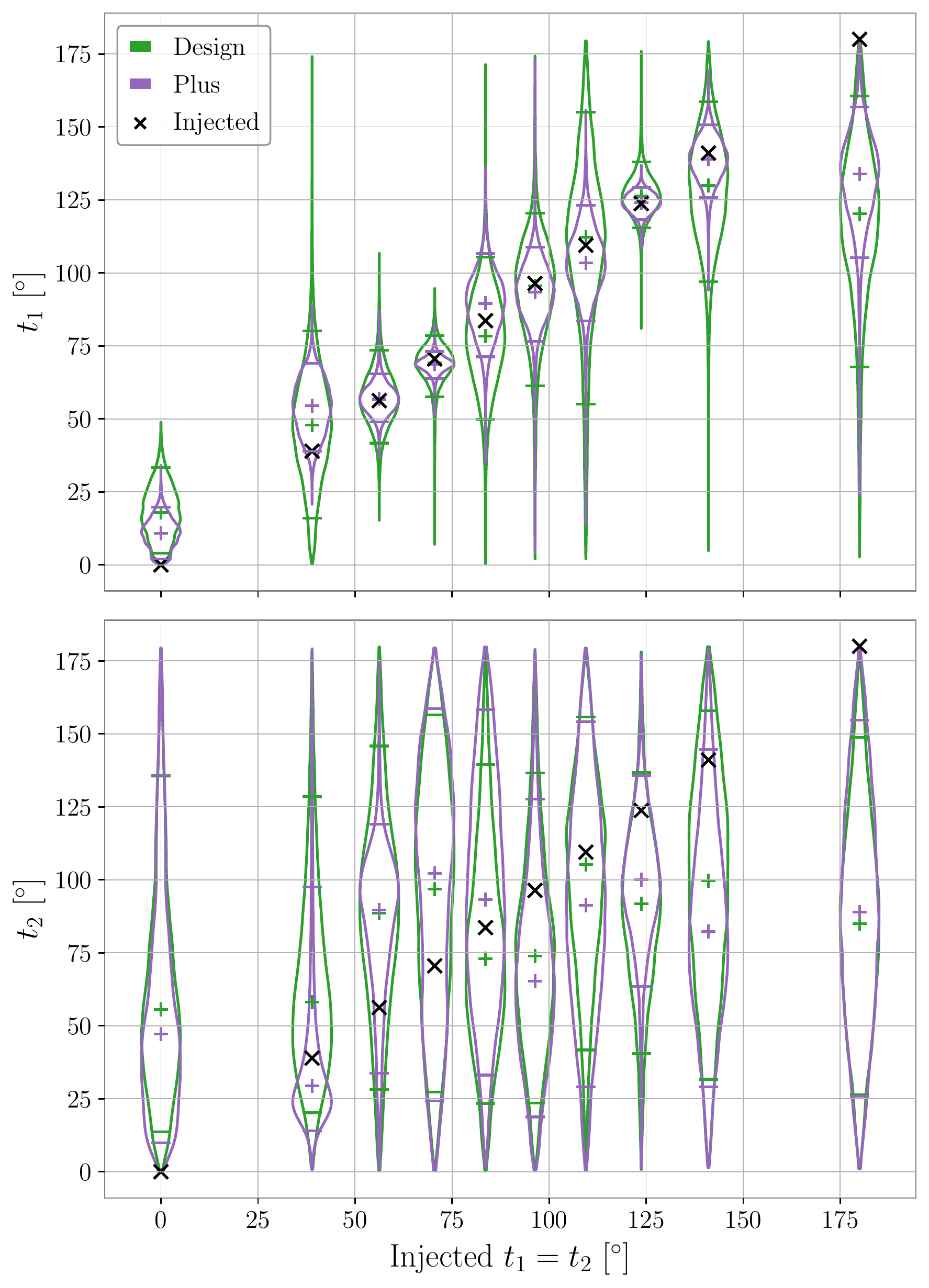}
\caption{Violin posterior plots of the two spin tilts, $t_1$ and $t_2$, for the spin tilt 1G+2G injection grid. The posteriors are arranged by their injected $t_1$, given on the horizontal axis. The green/purple pluses are the posterior medians, and the black crosses are the injected values. Horizontal bars mark the 5th and 95th percentiles of the posteriors, so that they enclose the 90\% CI. {\it Top:} Violin plots for $t_1$, showing the injections with equal component tilts, $t_1=t_2$.  {\it Bottom:} Violin plots for $t_2$, showing the same subset of the injection set. \label{fig:1G2G_violin}}
\end{figure}

Constraints on the two spin tilts are shown in Figure \ref{fig:1G2G_violin} for a subset of the injections with equal spin tilts $t_1=t_2$, which are representative of the wider set of results. For the aligned injections, we are able to recover the correct alignment at both sensitivities. The posteriors necessarily tend to zero at the prior boundaries because of the prior used for the spin tilts. However, because this prior places more support on $t_{1,2}\sim 90^\circ$, the fact that the posteriors have shifted towards $t_1\sim 0^\circ$ indicates strong support for an aligned primary spin. For the misaligned cases, the primary tilt is typically well-resolved with both networks, and the posteriors peak near the injected values. Plus sensitivity yields narrower constraints on $t_1$ compared to Design for both aligned and misaligned systems.

The case of anti-aligned spins appears to be the most challenging configuration, as it is here where we find broader posteriors with comparatively large offsets from the injected values. The weak tilt constraints for this case are another consequence of the shorter inspirals of these types of signals. 

Similar to the secondary magnitudes, the secondary spin tilts $t_2$ are often unconstrained, and their posteriors show little deviation from the prior in most cases.

\begin{figure}[t!]
\epsscale{1.15}
\plotone{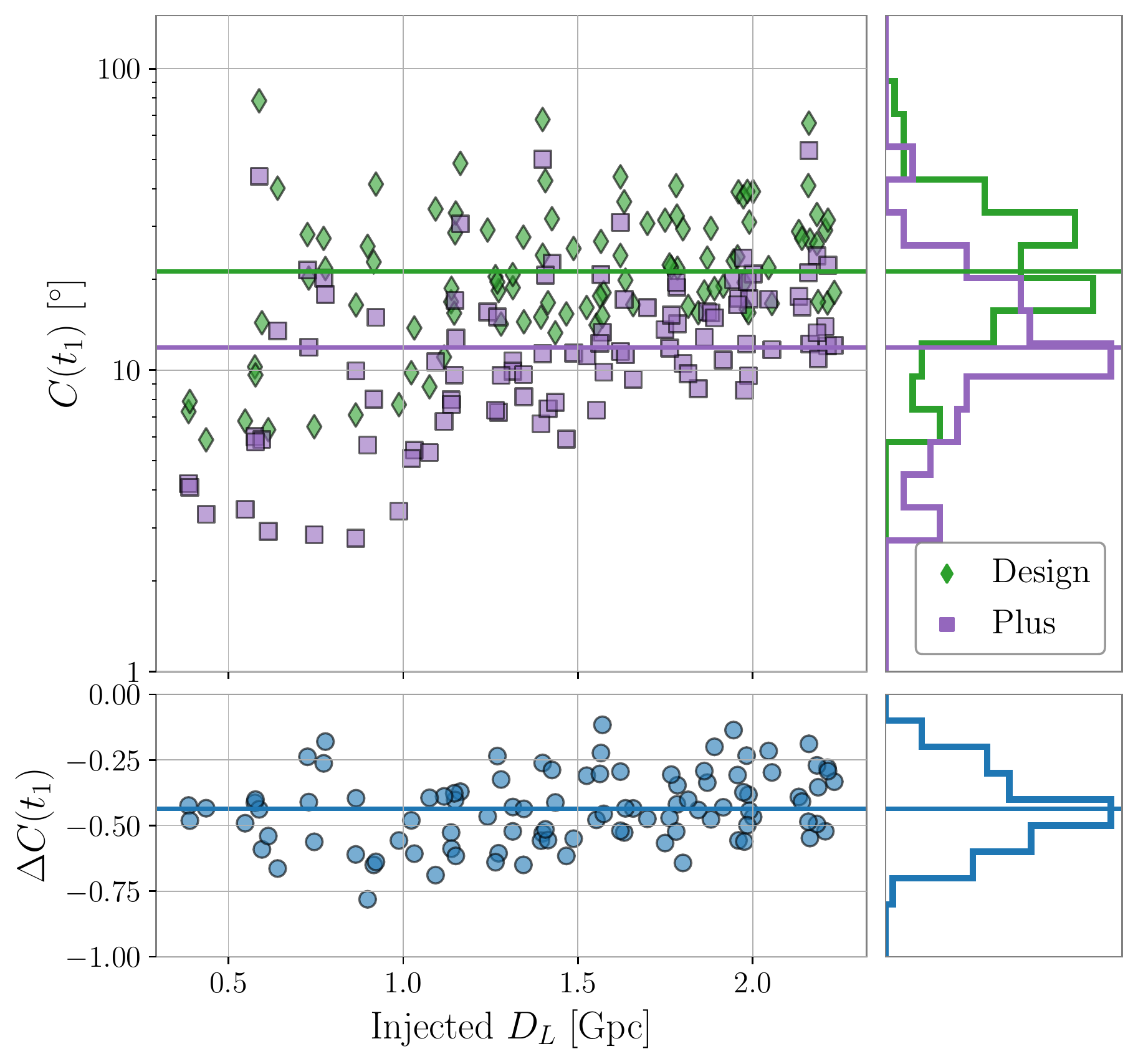}
\caption{Recovery of the primary tilt, $t_1$, as a function of the luminosity distance, $D_L$, for the spin tilt 1G+2G injection grid. The solid horizontal lines indicate the median of the quantity on the vertical axis. {\it Top:} Cost functions, $C(t_1)$, calculated for each posterior, with a histogram showing the distributions for the two networks. {\it Bottom:} Relative change in cost, $\Delta C(t_1)$, between Design and Plus sensitivity for each injected signal. \label{fig:1G2G_t1}}
\end{figure}

The full set of $t_1$ posteriors are represented using their cost functions, as shown in Figure \ref{fig:1G2G_t1}, where we find that $C(t_1)$ is of order 40\% smaller at Plus sensitivity compared to Design, which is similar to the $a_1$ constraints. We also note that there are several injections with $D_L<1$ Gpc that have their primary tilts constrained at the level of $C(t_1)\sim 3-5^\circ$, representing pristine resolution of the tilts. Additionally, 38\% of the injections have their primary tilt constrained at the level of $C(t_1)<10^\circ$, compared to just 11\% with Design. Therefore, the Plus network improves the rate of events with these exceptionally well-constrained spins by about a factor of three.  


\begin{figure}[t!]
\epsscale{1.15}
\plotone{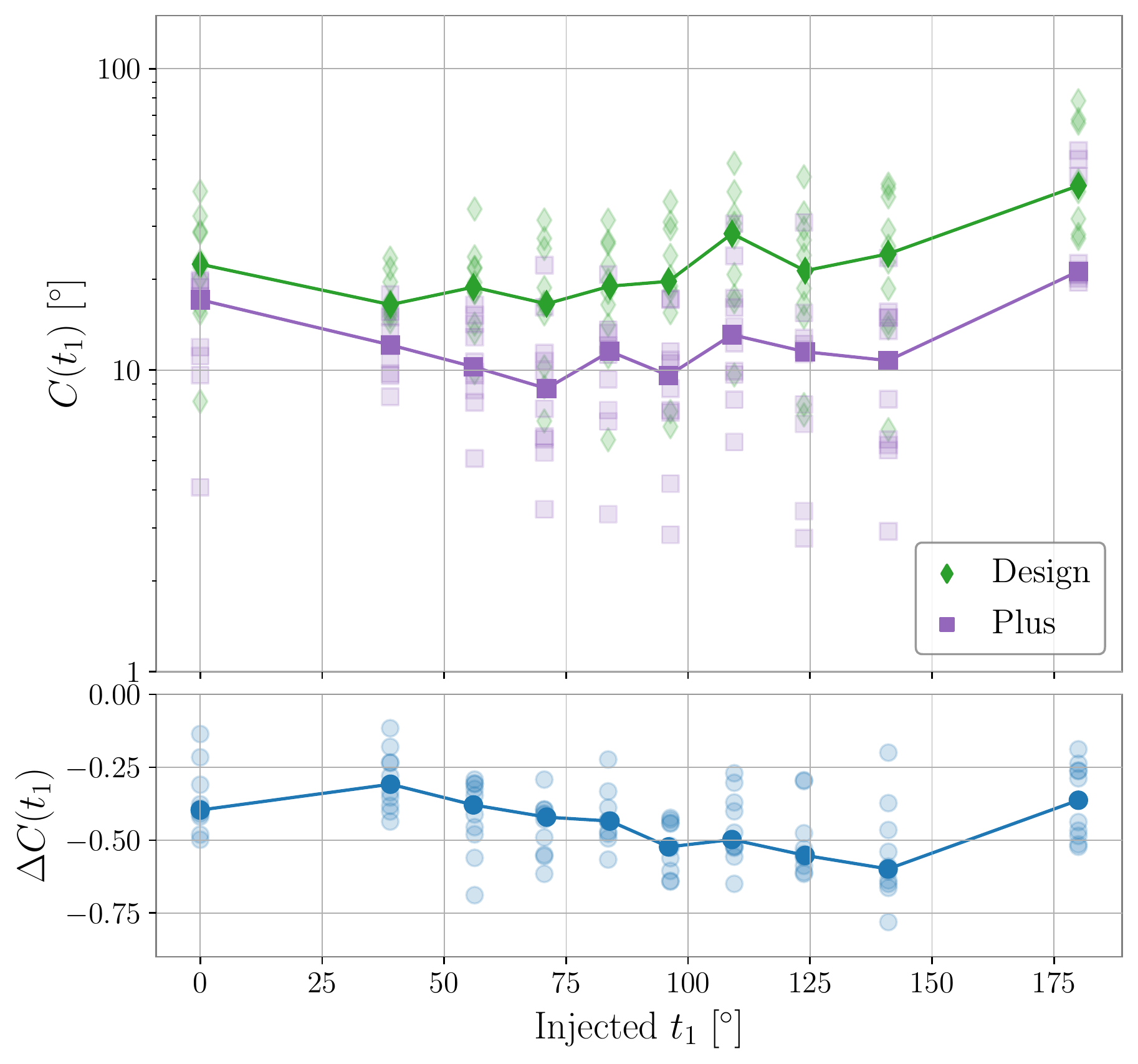}
\caption{Recovery of the primary tilt, $t_1$, as a function of the injected value of $t_1$ for the spin tilt 1G+2G injection grid. Transparent points correspond to individual posteriors. For each set of injections with the same value of $t_1$, we indicate the median of the quantity on the vertical axis with an opaque point. {\it Top:} Cost functions, $C(t_1)$, calculated for each posterior. {\it Bottom:} Relative change in cost, $\Delta C(t_1)$, between Design and Plus sensitivity for each injected signal. \label{fig:1G2G_t1_t1}}
\end{figure}

The relation between the tilt constraints and the injected values is shown in Figure \ref{fig:1G2G_t1_t1}, where we see the tilts are typically better measured when they are close to orthogonal. At Plus sensitivity, the injections with a misaligned primary tilt have median $C(t_1)\sim 10^\circ$, and for the aligned and anti-aligned injections the cost is about twice this amount. Similar to $a_1$, the largest improvements in the measurement of $t_1$ occur for signals with $90^\circ<t_1< 150^\circ$, where $\Delta C(t_1)$ reaches a minimum.


\subsection{Spin tilt 2G+2G grid}

\begin{figure}[t!]
\epsscale{1.15}
\plotone{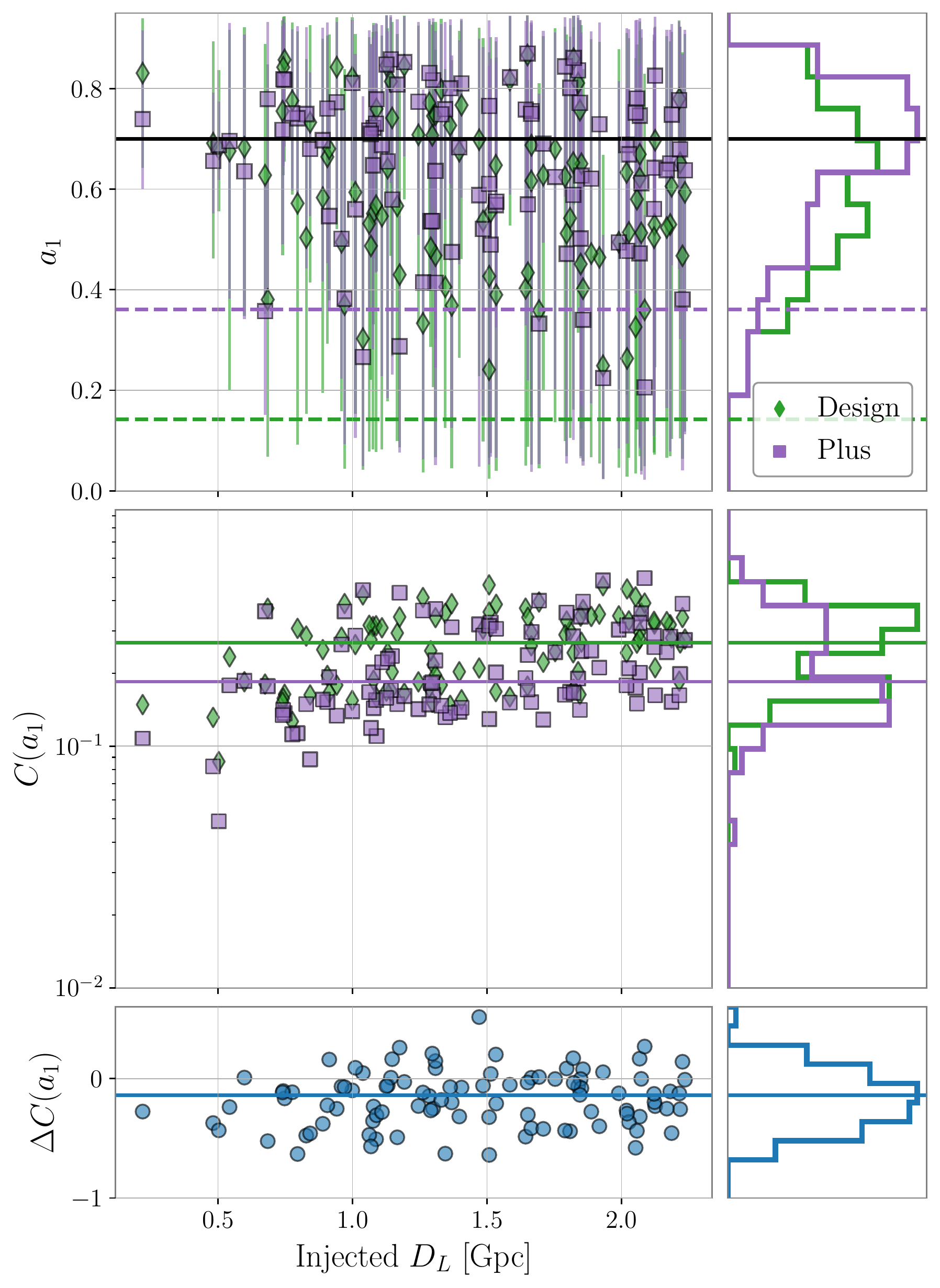}
\caption{Recovery of the primary spin magnitude, $a_1$, as a function of luminosity distance, $D_L$, for the spin tilt 2G+2G injection grid. {\it Top:} Posterior medians (squares/diamonds) with errorbars indicating the extent of the 90\% CIs (vertical lines). The histogram shows the distributions of the posterior medians for the Design and Plus networks. The injected value $a_1=0.7$ is indicated by the solid horizontal line. and the dashed horizontal lines show the median 5th percentiles of the CIs. {\it Middle:} Cost functions, $C(a_1)$, calculated for each $a_1$ posterior, with histograms showing the distributions for the two networks. The solid horizontal lines are the medians of the distributions. {\it Bottom:} Relative change in cost, $\Delta C(a_1)$, between Design and Plus sensitivity for each injected signal. A horizontal line marks the median. \label{fig:2G2G_a1}}
\end{figure}

\begin{figure}[t!]
\epsscale{1.15}
\plotone{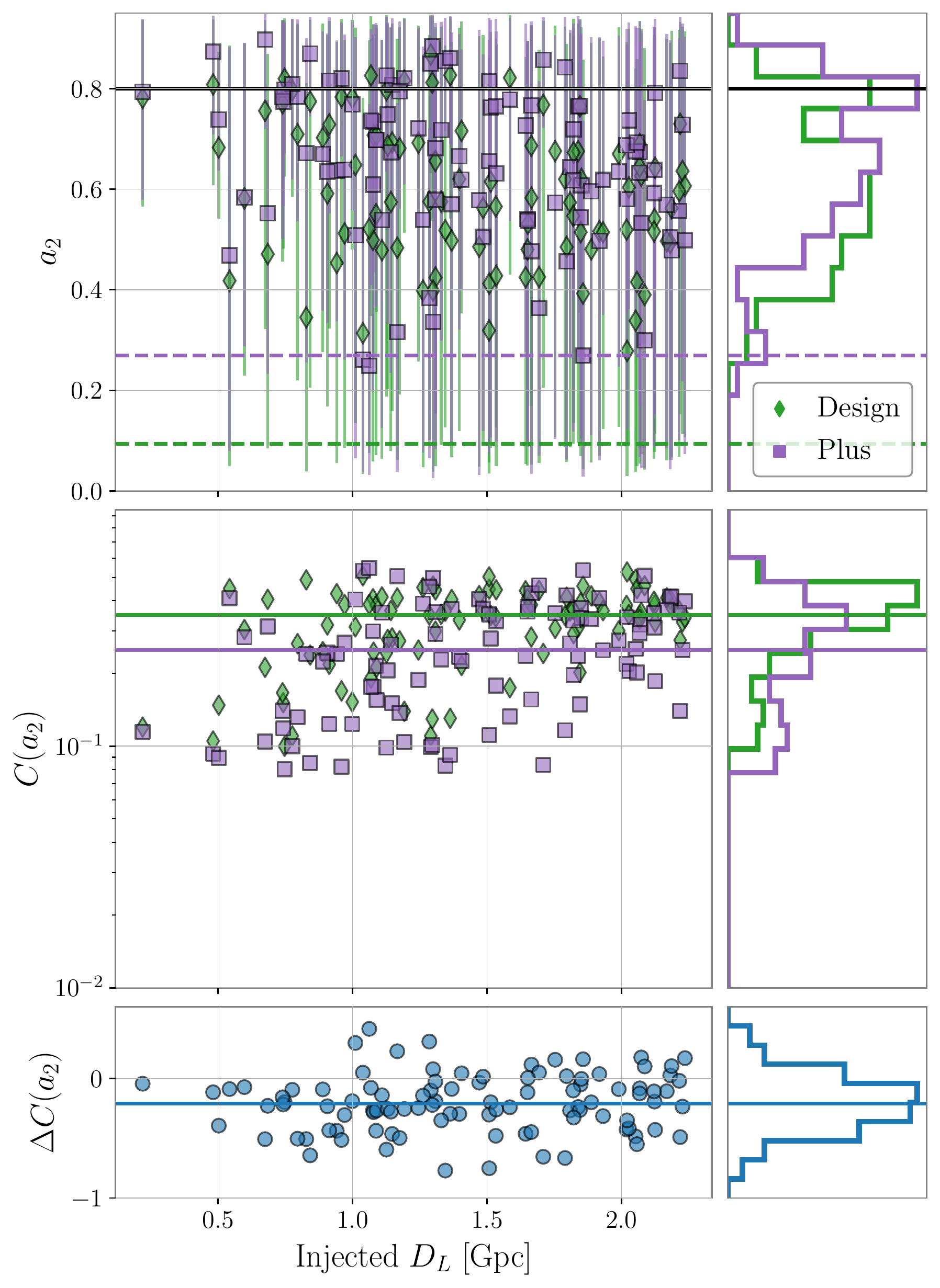}
\caption{Same set of plots as in Figure \ref{fig:2G2G_a1}, now for the secondary spin magnitude, $a_2$, where the injected value is 0.8. \label{fig:2G2G_a2}}
\end{figure}

We have seen that spin recovery is more effective when the masses are unequal, and that larger spins are better constrained. Figures \ref{fig:2G2G_a1}--\ref{fig:2G2G_a2} show the constraints on the two spin magnitudes $a_1$ and $a_2$ for systems consistent with a 2G+2G hierarchical merger, where the masses are comparable and each has a large spin. The median values of the 5th posterior percentiles are shown to ascertain the typical level of support for a high primary or secondary spin. Starting with the primary spin, the median 5th percentiles for Design and Plus sensitivity do not suggest particularly strong evidence for a large primary spin, especially when compared to the 1G+2G grid. Furthermore, at Plus sensitivity we find that 29\% of the injections recover $a_1>0.5$ with 95\% credibility, compared to 50\% in the 1G+2G grid for the same network. The $a_2$ constraints similarly do not suggest a consistent preference for large secondary spin, and only 26\% of the injections have constrained $a_2>0.5$ at the 95\% credible level with the Plus network.

\begin{figure}[t!]
\epsscale{1.15}
\plotone{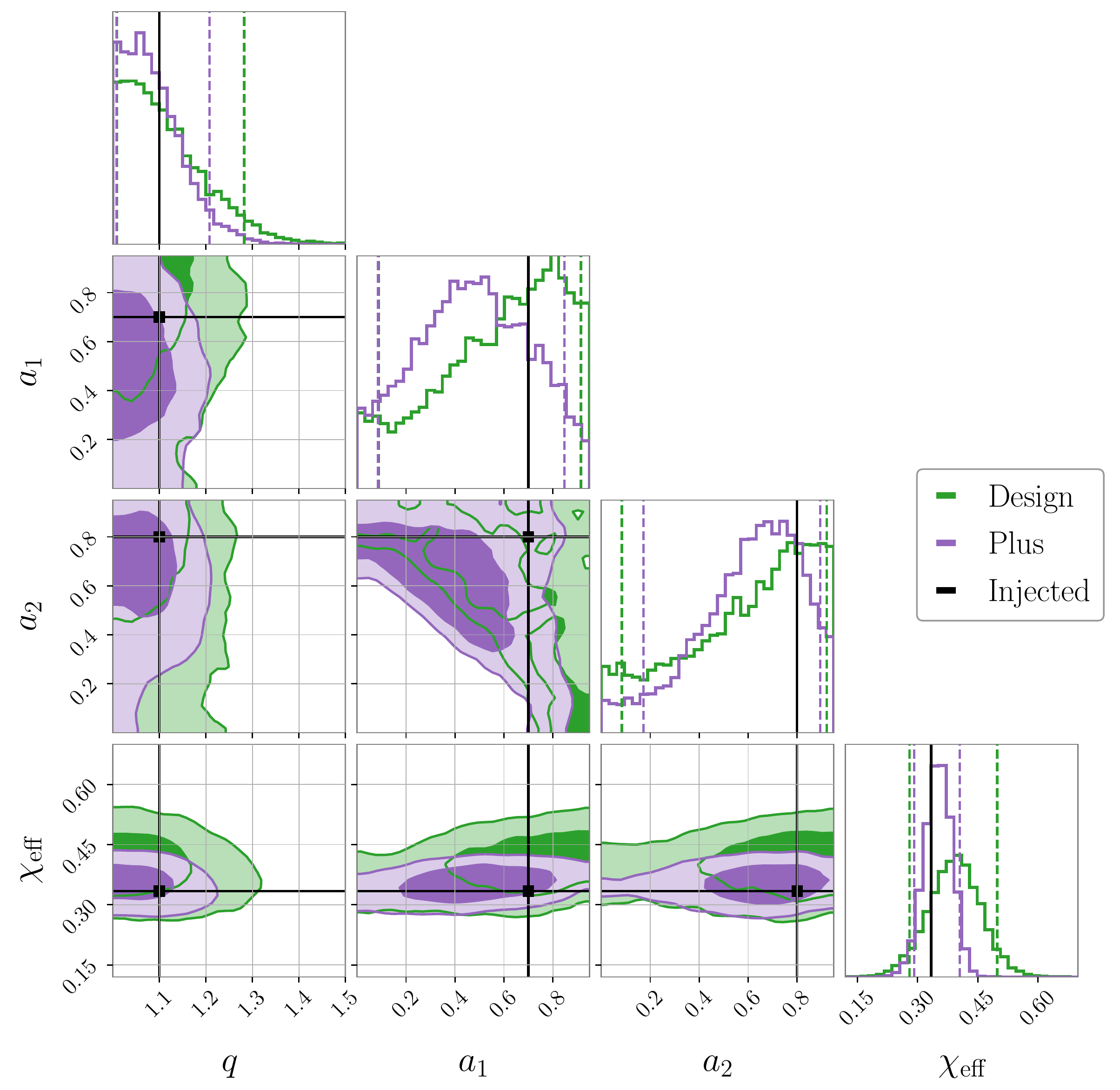}
\caption{Corner plot showing posteriors for an injection from the 2G+2G grid with $\Delta C(a_1)=0.17$, demonstrating the $a_1$--$a_2$ degeneracy. The contours represent the 50\% (darker shading) and 90\% (lighter shading) credible regions, and the vertical lines in the histograms show the 5th and 95th percentiles for the one-dimensional posteriors. The injected values are marked in red. \label{fig:2G2G_corner}}
\end{figure}

Owing to its larger magnitude, the median reduction in $C(a_2)$ is roughly 20\% between the detector two networks, which is an improvement over the 1G+2G grid. Conversely, $a_1$ is not measured as well. The medians of $C(a_1)$ at Design and Plus sensitivity are roughly 1.5 times their respective values for the 1G+2G grid. This is in part because of a spin-spin degeneracy between $a_1$ and $a_2$ that is more significant for (near) equal-mass systems. For the 2G+2G injections, Plus sensitivity tends to yield less support for small secondary spins compared to Design. If this is not accompanied with reduced posterior support for $q=1$, then this leads to a decrease in support for a large $a_1$ so that $\chi_{\rm eff}$ is conserved. As a result, the recovered $a_1$ is a slight underestimate of the injected value. An example of this degeneracy playing out is shown in Figure \ref{fig:2G2G_corner}, in which the Plus network does not exclude $q=1$, but instead slightly increases support for equal masses. In this particular case, the shifting in the $a_1$ posterior results in a higher cost with the Plus network than Design ($\Delta C(a_1)>0$). This degeneracy was not as important in the 1G+2G grid because it is easier to resolve the more asymmetric mass ratios of those injections, which helps to disentangle the two spin terms in $\chi_{\rm eff}$. 

\begin{figure}[t!]
\epsscale{1.15}
\plotone{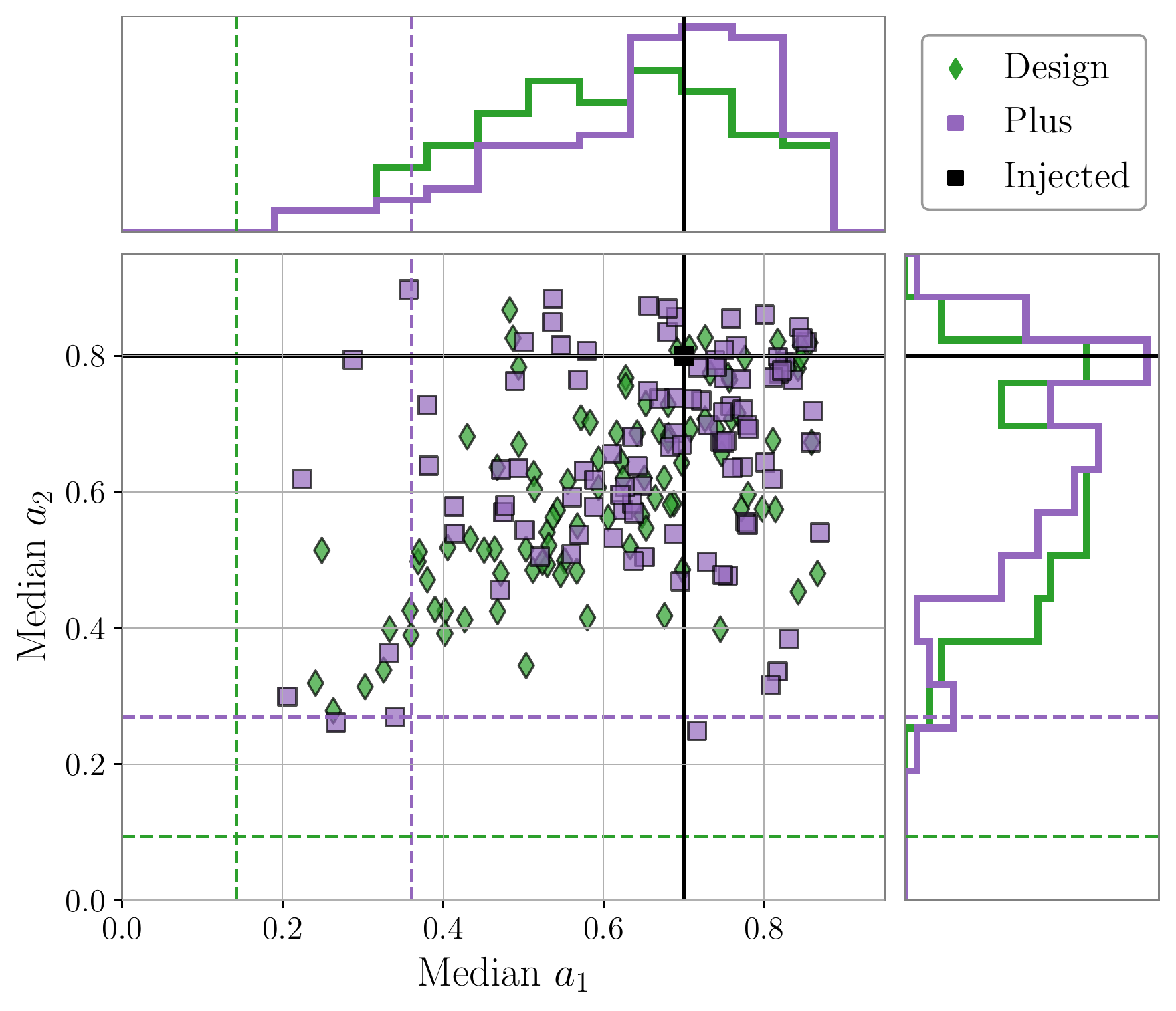}
\caption{Median posterior estimates for the primary spin magnitudes, $a_1$ and $a_2$, for the 2G+2G injection grid, with histograms showing how they are distributed. The dashed horizontal and vertical lines are the median values of the 5th posterior percentiles. The injected values are indicated by the black square. \label{fig:2G2G_medians}}
\end{figure}

\begin{figure}[t!]
\epsscale{1.15}
\plotone{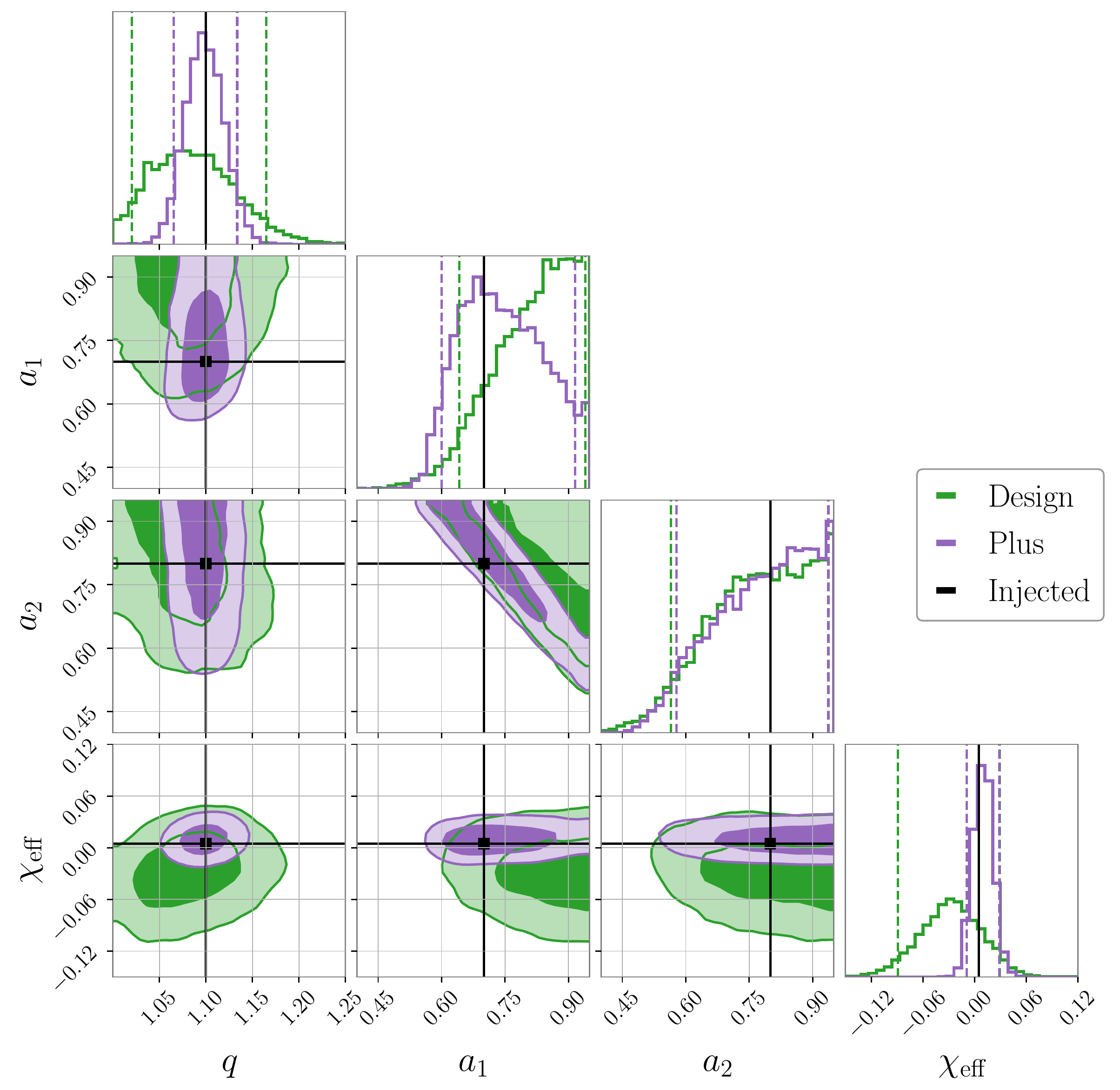}
\caption{Corner plot showing posteriors for an injection with nearly zero effective aligned spin, $\chi_{\rm eff}$, from the 2G+2G grid, showing a case where both spin magnitudes are accurately measured. Contours and colours are the same as in Figure \ref{fig:2G2G_corner}. \label{fig:2G2G_corner2}}
\end{figure}

An important question is whether we can accurately recover the large spins of both components simultaneously, as this is a characteristic signature of 2G+2G mergers. In Figure \ref{fig:2G2G_medians}, we plot the median posterior estimates of $a_1$ and $a_2$ on the same plane. For 17\% of the injections, both magnitudes are constrained above 0.5 with 95\% credibility at Plus sensitivity, compared to 5\% at Design. We show posteriors for one such injection in Figure \ref{fig:2G2G_corner2}. We see that the Plus network resolves the slightly unequal mass ratio, and hence the inference does not suffer as much from the $a_1$--$a_2$ degeneracy as discussed earlier. The shifting in the posterior away from $q=1$ also lets us rule out most of the negative $\chi_{\rm eff}$ region that fell into the 90\% CI at Design sensitivity.

\begin{figure}[t!]
\epsscale{1.15}
\plotone{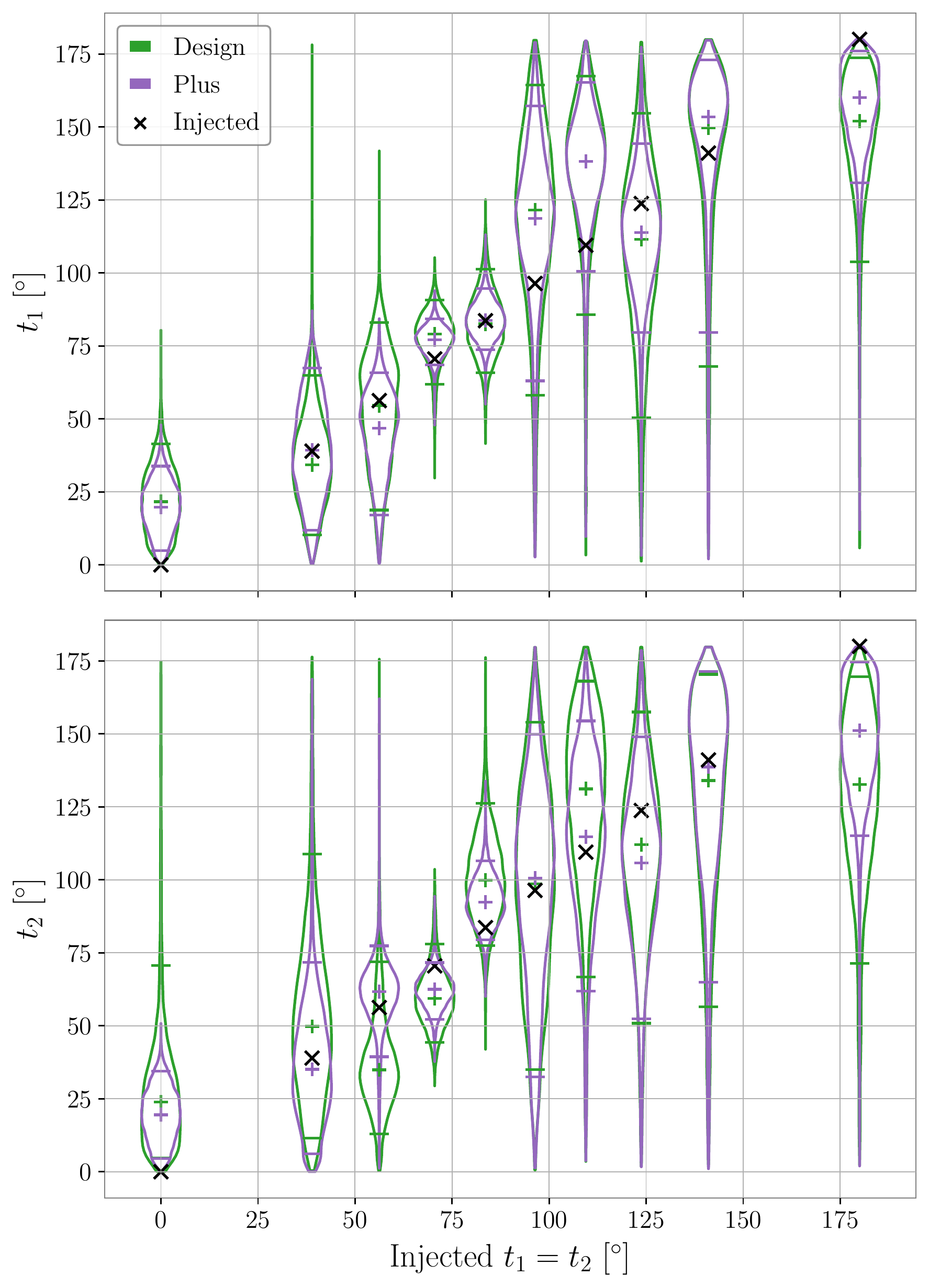}
\caption{Violin posterior plots of the two spin tilts, $t_1$ and $t_2$, for the spin tilt 2G+2G injection grid. Shown here are injections with equal tilts, $t_1=t_2$. The posteriors are arranged by their injected $t_1$, given on the horizontal axis. The green/purple pluses are the posterior medians, and the black crosses are the injected values. Horizontal bars mark the 5th and 95th percentiles of the posteriors, so that they enclose the 90\% CI. {\it Top:} Violin plots for $t_1$. {\it Bottom:} Violin plots for $t_2$. \label{fig:2G2G_violin}}
\end{figure}

\begin{figure}[t!]
\epsscale{1.15}
\plotone{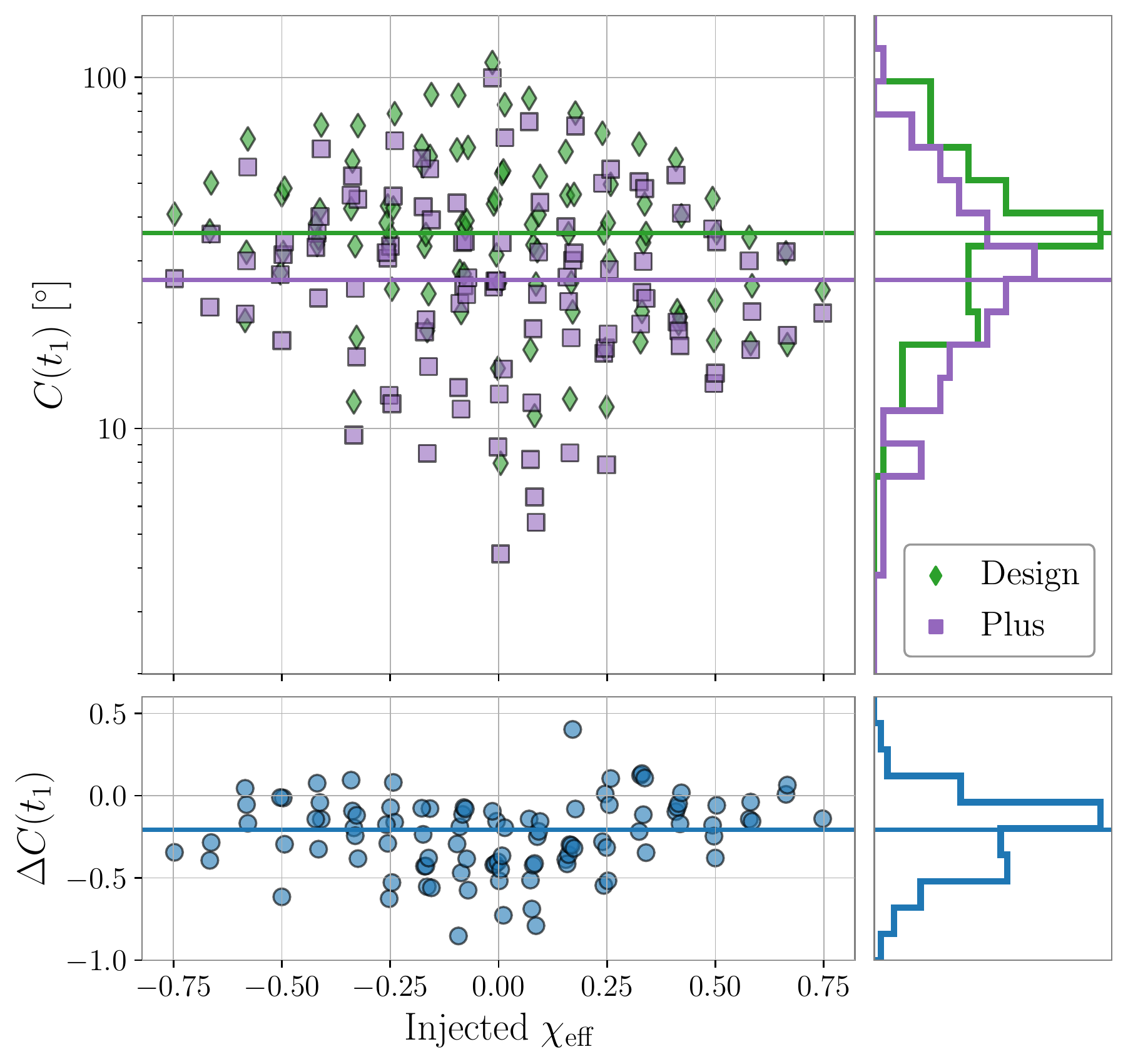}
\caption{Recovery of the primary tilt, $t_1$, as a function of effective aligned spin, $\chi_{\rm eff}$, for the spin tilt 2G+2G injection grid. The solid horizontal lines indicate the median of the quantity on the vertical axis. {\it Top:} Cost functions, $C(t_1)$, calculated for each posterior, with a histogram showing the distributions for the two networks. {\it Bottom:} Relative change in cost, $\Delta C(t_1)$, between Design and Plus sensitivity for each injected signal. \label{fig:2G2G_t_1}}
\end{figure}

Next we investigate the constraints on the two spin tilts, shown in Figure \ref{fig:2G2G_violin}.  Similar to the spin magnitudes, we find that constraints on the primary tilt are typically not as well-measured as in the 1G+2G grid. We also notice that there are larger offsets in some of the estimates, especially when the injected values of the tilts are greater than $90^\circ$. The secondary tilt is better measured than in the 1G+2G grid, but constraints tend to broaden again as the tilts become anti-aligned, where we also see that there is less improvement between the two networks.

Because both of the tilts are varied in this grid, and both components of the binary contribute greatly to the waveform due to their comparable masses and large spins, it is difficult to identify any correlations by plotting the cost function over the injected tilts as for the 1G+2G grid. Rather, we find it more convenient to show how the constraints vary with the effective aligned spin $\chi_{\rm eff}$, shown in Figure \ref{fig:2G2G_t_1}. We only include the primary tilt since the overall distribution for $t_2$ appears similar and conveys the same trend that we describe here. We see that $\Delta C(t_1)$ tends towards lower values for systems with negligible aligned spin, $\chi_{\rm eff}\sim 0$. The spin parameters are most degenerate around zero $\chi_{\rm eff}$, since there are a greater number of combinations of $a_{1,2},t_{1,2}$ that yield zero $\chi_{\rm eff}$ as opposed to a large positive or negative $\chi_{\rm eff}$, especially when $q\sim 1$ (this is clear from the shape of the $\chi_{\rm eff}$ distribution assuming isotropic spin priors, see e.g.~\citet{Callister:2021gxf}). The improved sensitivity of the Plus network clearly helps to mitigate this degeneracy, in part by resolving the slight mass inequality which leads to a more tightly constrained $\chi_{\rm eff}$ compared to Design.

In contrast to the 1G+2G grid, only 9\% of the injections have $t_1$ measured with $C(t_1)<10^\circ$ at Plus sensitivity, roughly four times lower than for the 1G+2G grid, and almost none (1\%) have $t_1$ constrained at this level with Design sensitivity. The same is true of $t_2$, where we find similar percentages. The median reduction in cost between the two detector networks is on the order of 20\% for both spin tilts, which for $t_1$ is a factor of two smaller than in the 1G+2G grid. We thus conclude that measuring spins is significantly more challenging for 2G+2G mergers, as the lower mass ratio and similar spin magnitudes suppresses the accuracy of spin recovery both in absolute terms and in terms of the relative improvement between the Design and Plus networks.


\subsection{Population study}

In this section, we simulate 200 events drawn from the ``power law + peak'' mass and ``default'' spin population models introduced in Section \ref{sec:grids}. We examine how well the spins can be constrained in this injection set and to what degree the behaviours observed in the previous grids still apply when simulating a more realistic population of events.

\begin{figure}[t!]
\epsscale{1.15}
\plotone{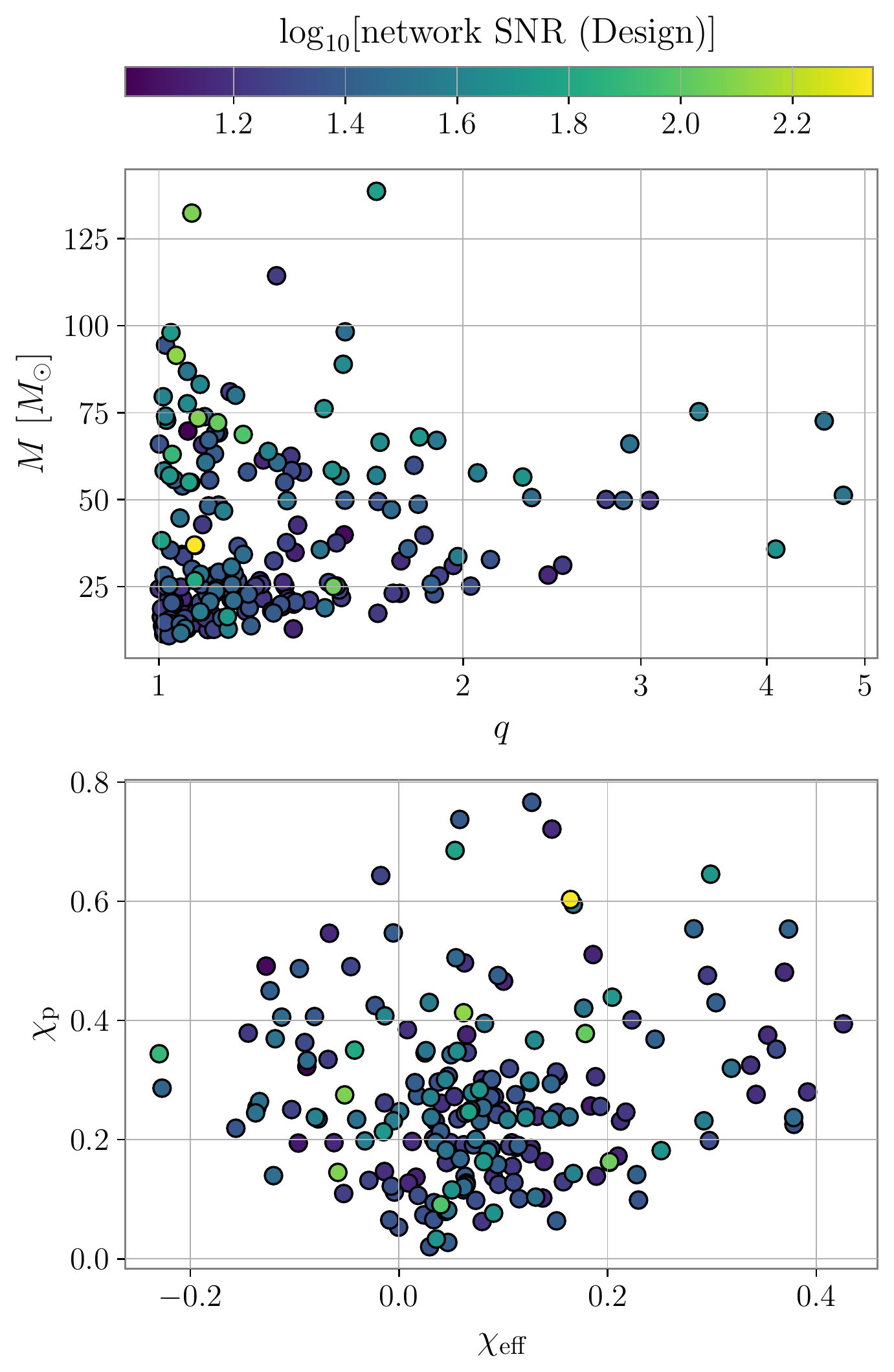}
\caption{Source parameter distributions of our population injections. Each injection is represented by a point in the $M$--$q$, where $M$ is the source-frame total mass and $q$ is the mass ratio, and a point in the $\chi_{\rm eff}$--$\chi_{\rm p}$ plane, where $\chi_{\rm eff}$ and $\chi_{\rm p}$ are the effective aligned and precession spins. The points are coloured by the Design network SNR. \label{fig:pop_hist}}
\end{figure}

The intrinsic parameters of the resulting injection set are shown in Figure \ref{fig:pop_hist}. The set contains total (source-frame) masses ranging from $11\,M_\odot$ to $139\,M_\odot$. The vast majority (93\% of the injections) have mass ratios between $1\leq q\leq 2$, with about a dozen $q>2$ sources. The $\chi_{\rm eff}$ distribution is skewed towards positive effective spins due to a subpopulation with preferentially aligned spin. Systems with both spins tilted beyond $90^\circ$ are rare, hence there are relatively few samples with negative $\chi_{\rm eff}$. Nonetheless, the injections incorporate varying degrees of precession, as shown in Figure \ref{fig:pop_hist}. Most of the injections (95\%) have network SNR between 8 and 60 at Design sensitivity, with just a few high-SNR injections above 100. 

\begin{figure}[t!]
\epsscale{1.15}
\plotone{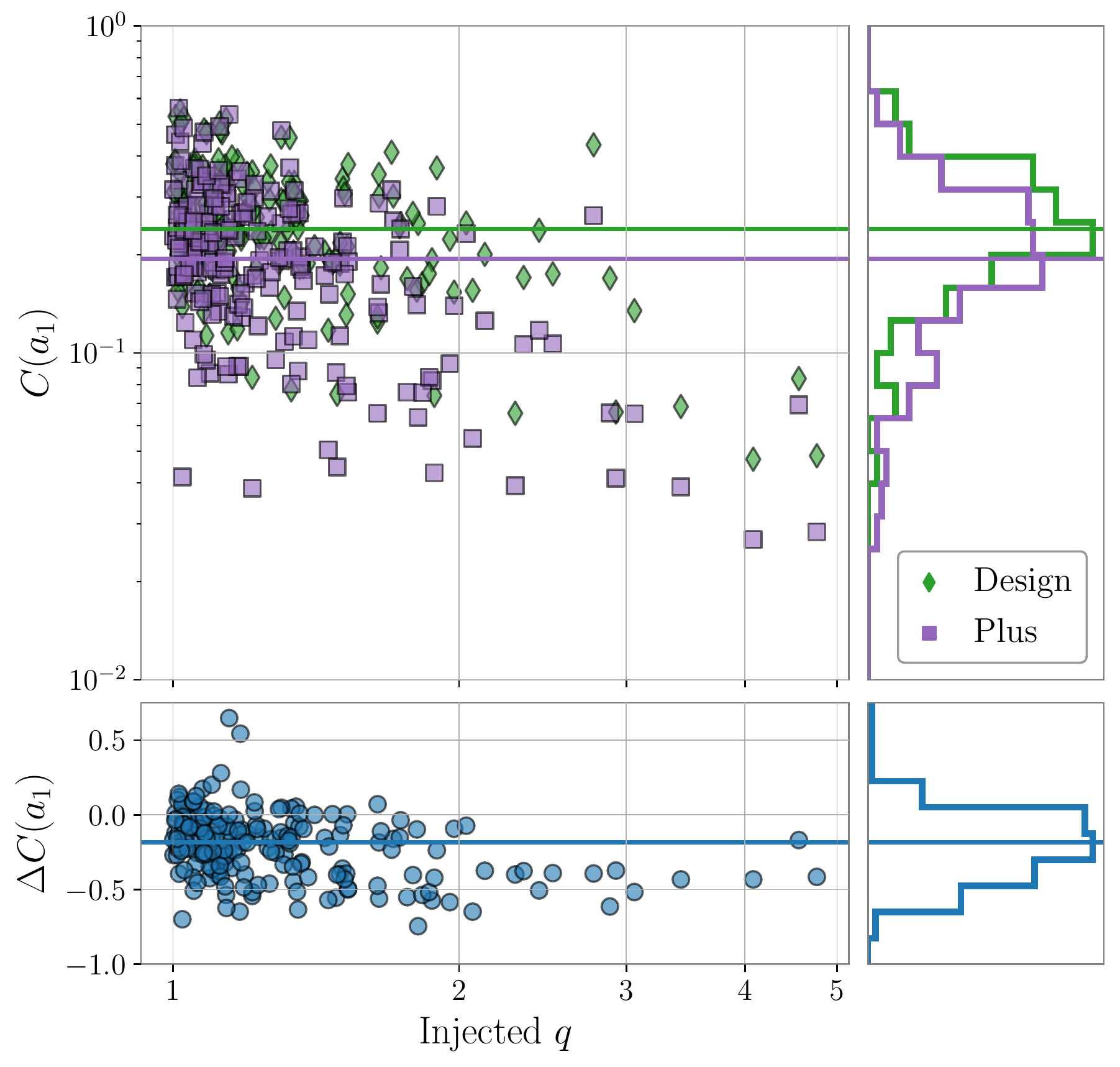}
\caption{Recovery of the primary spin magnitude, $a_1$, plotted against the mass ratio, $q$, for the population injection set. The solid horizontal lines indicate the median of the quantity on the vertical axis. {\it Top:} Cost functions $C(a_1)$, calculated for each posterior, with a histogram showing the distributions for the two networks. {\it Bottom:} Relative change in cost, $\Delta C(a_1)$, between Design and Plus sensitivity for each injected signal. \label{fig:pop_a_1}}
\end{figure}

\begin{figure}[t!]
\epsscale{1.15}
\plotone{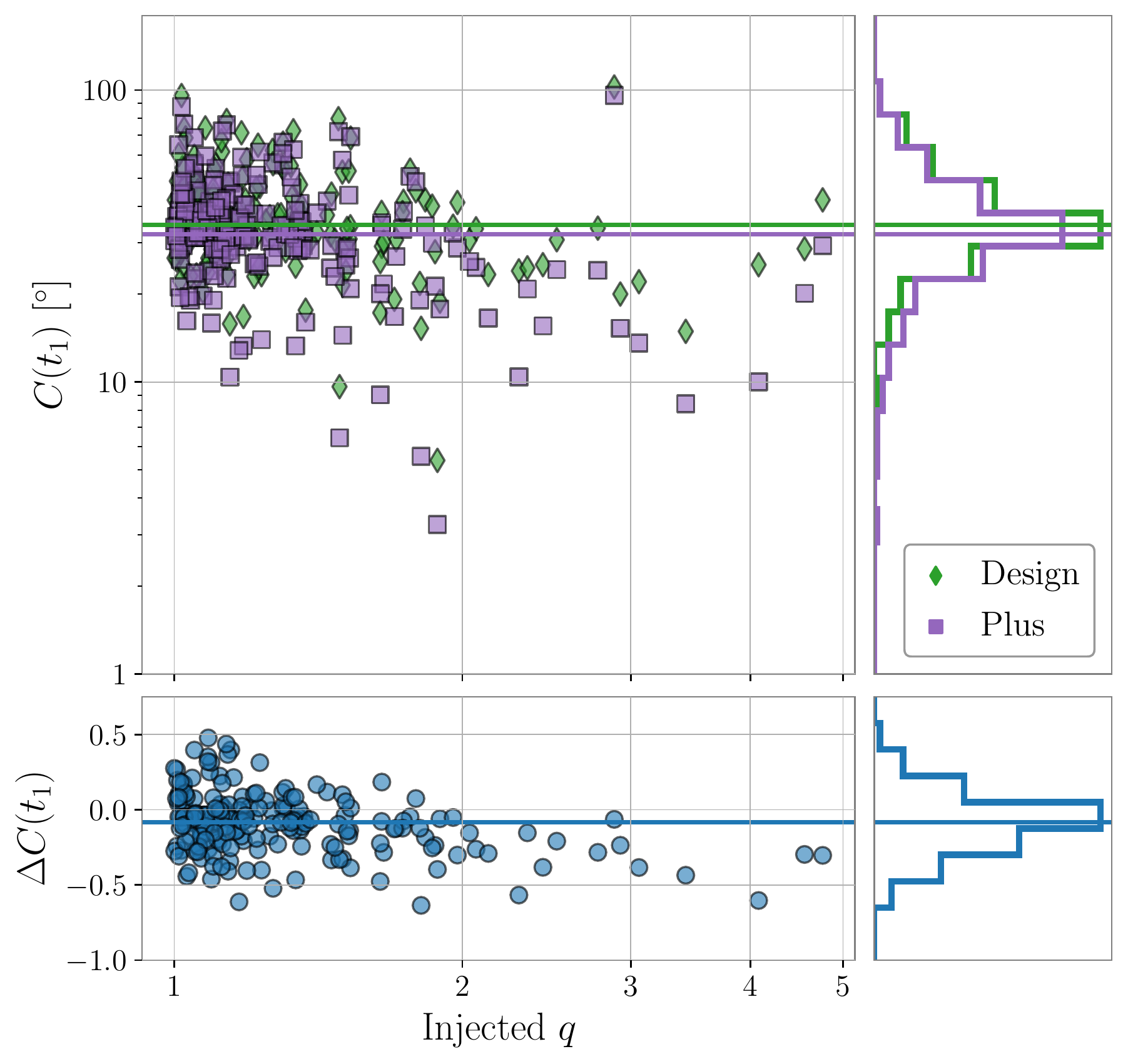}
\caption{Same set of plots as Figure \ref{fig:pop_a_1}, now for the primary tilt, $t_1$. \label{fig:pop_t_1}}
\end{figure}

The population's tendency toward equal masses and low spins will hinder spin estimation. Indeed, a significant fraction of the injections return uninformative posteriors for the primary spin, as seen in Figures \ref{fig:pop_a_1}--\ref{fig:pop_t_1} where we show the distribution of constraints on $a_1$ and $t_1$. The median reduction in cost for $a_1$ is on the order of 20\%, which is comparable to the 2G+2G grid. For $t_1$, the median reduction is $-0.08$, so there is typically minimal change in the primary tilt constraints between the two networks. Nonetheless, we start to find more informative spin constraints as the mass ratio increases, which is consistent with our previous discussions thus far. When including only the injections with $q>2$, the Plus network consistently performs better than Design ($\Delta C<0$) in terms of constraining the primary spin, and the median reduction in cost rises to 40\% and 25\% for $a_1$ and $t_1$, respectively.

As the two spin magnitudes are drawn from the same distribution, some injections will have $a_2$ greater than or comparable to $a_1$, which can allow us to place stronger constraints on the secondary spin (e.g.~unlike the 1G+2G grid where $a_2$ is small). The median cost functions for both $a_1$ and $a_2$ are similar to the 2G+2G results ($\sim 20\%$ reduction in cost), whereas for $t_2$ there is essentially no change between the two networks, even for injections with $q>2$.

\begin{figure*}[t!]
\epsscale{1}
\plotone{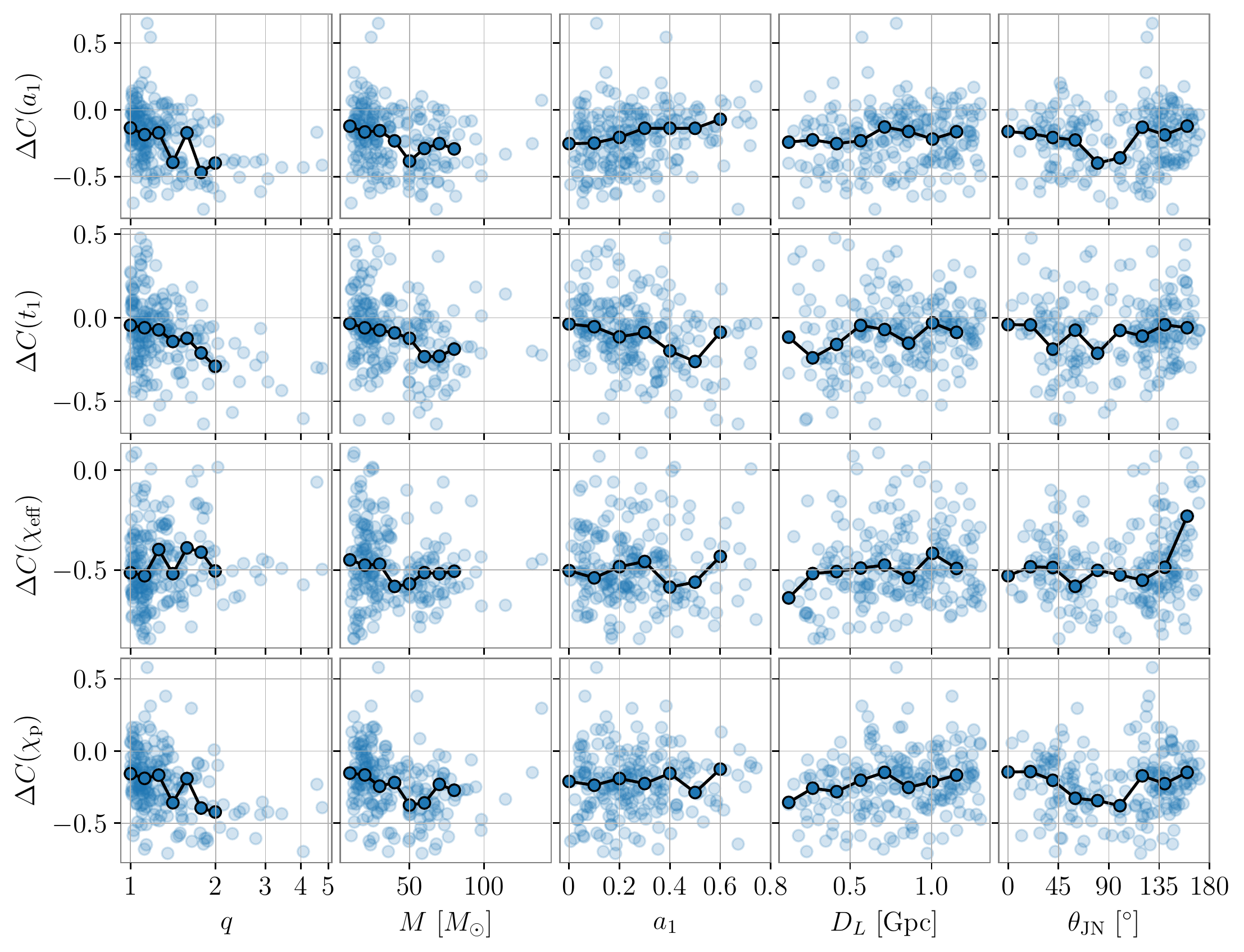}
\caption{Constraints on the primary spin magnitude, $a_1$, tilt, $t_1$, effective aligned spin, $\chi_{\rm eff}$, and effective precession spin, $\chi_{\rm p}$, for the population set, as functions of $a_1$, the injected mass ratio, $q$, source-frame total mass, $M$, luminosity distance, $D_L$, and inclination of the total angular momentum relative to the line-of-sight, $\theta_{\rm JN}$. The cost functions for individual posteriors are shown as the transparent points. The injections are also binned by the source parameter on the horizontal axis. We calculate the median relative change in cost $\Delta C$ between Design and Plus for each bin, indicated with opaque points, which also mark the left-edges of the bins. The $q$-bins are logarithmically spaced, and the rest are linearly spaced. Some bins were adjusted to ensure each one contains no less than 10 injections. \label{fig:pop_grid}}
\end{figure*}

It is also interesting to look at how the constraints correlate with source parameters that were previously fixed in the earlier grids. To make the correlations clearer, we binned the injections by parameters of interest: the mass ratio, source-frame total mass, primary spin magnitude, luminosity distance, and inclination, and calculated the median of the relative change in the cost function $\Delta C$ for the injections in each bin. These results are shown in Figure \ref{fig:pop_grid}, where we show $\Delta C$ for the primary spin magnitude and tilt, and the two effective spin parameters.

The top two rows of Figure \ref{fig:pop_grid} show the constraints for the primary spin as a function of the chosen injected source properties. Constraints on the primary spin greatly improve with increasing mass ratio, as we expect. The total mass of the system will affect spin estimation in a variety of ways, being an important factor in determining the SNR and what fraction of the signal power comes from the inspiral versus the merger and ringdown. Lower-mass systems are inspiral-dominated, which is understood to enhance spin estimation. However, it is for the higher-mass systems, particularly those with $M\gtrsim 50\,M_\odot$, that Plus sensitivity recovers the primary spin even more accurately relative to Design. 

Plus sensitivity also recovers $a_1$ more accurately for low-spin systems, and gains in constraints gradually decrease as the primary spin increases until there is minimal difference between the two networks. This is a departure from the mass-spin grid, where larger spins led to greater improvement between the two networks, and so this trend does not appear robust against more realistic variations in the other intrinsic source properties.

The cost functions for the primary spin reach a minimum at edge-on inclinations ($\theta_{\rm JN}\sim 90^\circ$), as we expect. Gravitational radiation from edge-on systems can observed from above and below the orbital plane if it is precessing, which is known to improve spin estimation \citep{Vitale:2014mka}. We find that, in addition to the cost functions reaching a minimum for (near) edge-on systems, the largest improvements between Design and Plus are also found for such systems. This behaviour is present in the measurement of the primary spin, as well as the effective precession spin, $\chi_{\rm p}$, showing that stronger measurements of spin/precession are in part being driven by systems with edge-on orientation.

\begin{figure}[t!]
\epsscale{1.15}
\plotone{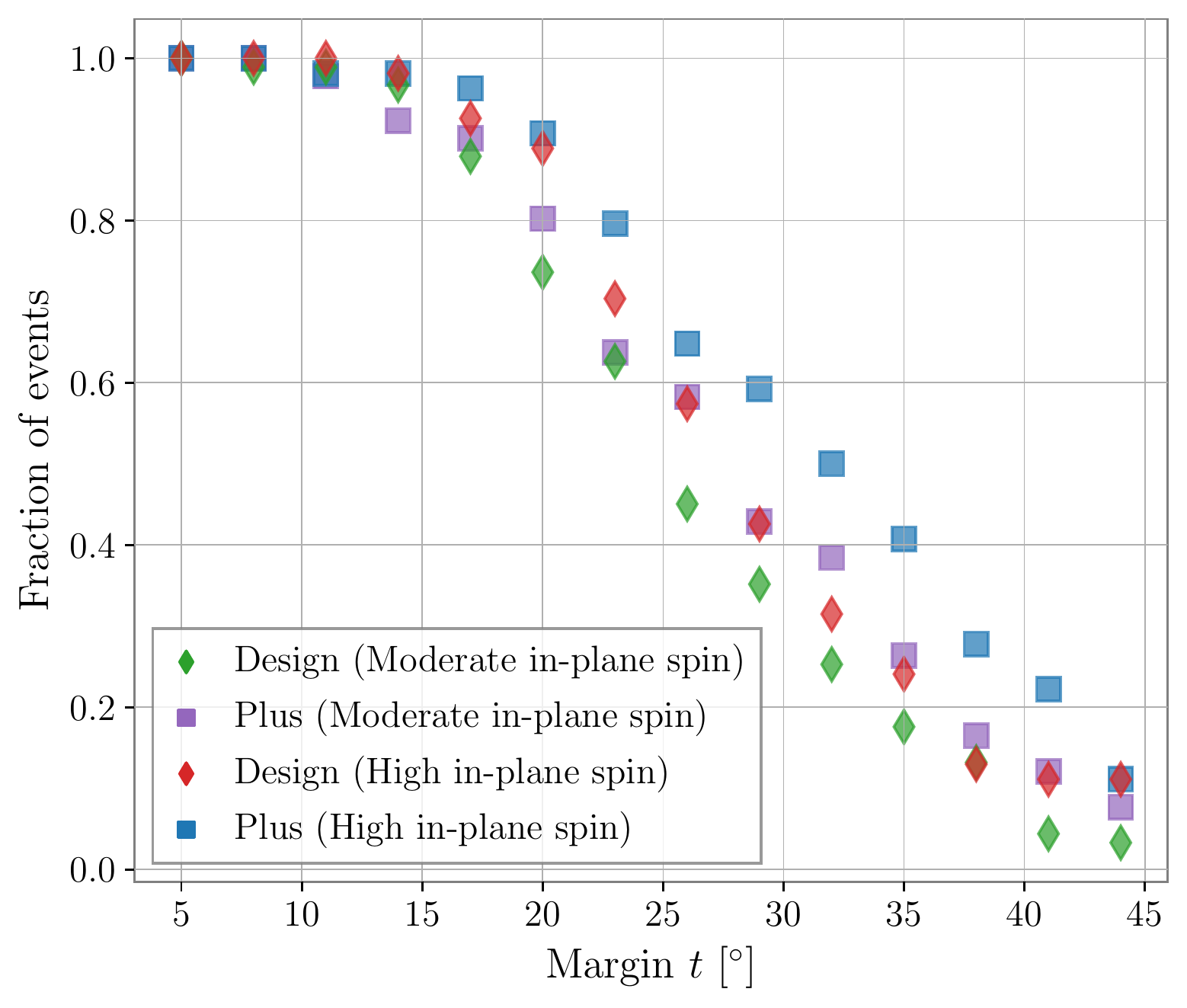}
\caption{Fraction of injections in two subsets of the population set where an aligned and anti-aligned primary spins are simultaneously excluded at 90\% credibility by some symmetric margin $t$, plotted on the horizontal axis. The colours and shapes distinguish the network and in-plane spin subset.  \label{fig:pop_rates}}
\end{figure}

We also include two more rows of plots for $\chi_{\rm eff}$ and $\chi_{\rm p}$, as these values are often reported parameter estimation studies. $\chi_{\rm eff}$ is consistently well-measured with a median reduction in cost of roughly 50\% between the two detector networks throughout the full range of masses and spin. The trends in $\chi_{\rm p}$ mirror the trends we observed for the primary spin.

Lastly, we study the rate at which we can confidently resolve misaligned spins in this injection set. In particular, we look at how often we can exclude an aligned ($t_1=0^\circ$) or anti-aligned ($t_1=180^\circ$) primary spin as a function of an increasingly aggressive criterion. First, we divide the population set into two subsets: those with ``moderate'' in-plane spin, $45^\circ\leq t_1\leq 75^\circ$ or  $105^\circ\leq t_1\leq 135^\circ$, and with ``high'' in-plane spin, $75^\circ\leq t_1\leq 105^\circ$. Given a margin $t$, we compute the fraction of injections in each subset for which the 5th percentile of the $t_1$ posterior is greater than $t$, and the 95th percentile is less than $180^\circ-t$. These fractions are shown for each detector network as a function of $t$ in Figure \ref{fig:pop_rates}. The horizontal axis stops before $t=45^\circ$ because this is the minimum allowable tilt in the moderate subset. We can see that the points at $t=35^\circ$ tell us that at Plus sensitivity, 40\% of the high in-plane spin injections have their primary tilt constrained within the range $35^\circ\leq t_1\leq 145^\circ$ at the 90\% credible level or higher. Considering the same subset at Design sensitivity, only $\sim 22\%$ have $t_1$ constrained in this same range and credibility. 

The fractions tend to $1$ for all the subsets as the margin becomes smaller, since we have chosen injections with $t_1$ well away from the prior boundaries and it is quite easy to exclude these small regions of the parameter space with low prior probability. As we increase the margin, we begin to exclude the less informative posteriors that span most of the prior range, and the different spin subsets and detector networks begin to diverge from each other. For margins $30^\circ<t<40^\circ$, we find that the Plus network can exclude an aligned or anti-aligned primary spin at $1.5-2$ times higher rates than Design. Very aggressive margins of around $45^\circ$ result in nearly all the injections being excluded, which is why the rates for Design and Plus converge again. We generally find better improvements between detector networks for the high in-plane spin subset.

\section{Summary}\label{sec:disc}


Spin misalignment and precession in detected BBHs are interesting potential indicators of the formation of these systems. The ability of near-future gravitational-wave detectors to accurately measure the signatures of off-axis spins will have important implications for the study of these formation channels. In order to understand what new spin information might be revealed by the expected improved sensitivity of these detectors, we have performed an extensive comparison of spin recovery between the Design and Plus three-detector networks using a large-scale injection campaign, targeting a wide range of source properties. In the first part of our study, we analyzed grids of injections in which we systematically varied pairs of parameter in order to determine how they affect spin estimation. Then, we analyzed an injection set sampled from astrophysical population models to see which trends from the injection grids applied to a more realistic set of events.


We found the Plus network yields the largest improvements in spin estimation over Design in systems with high mass ratios and high spin magnitudes, which we attributed to the breaking of degeneracy between the masses and spins due to the higher-order modes and precession. In particular, for systems with mass ratios $q\geq 2$ we found that the uncertainty in spin measurements is halved as a result of the improved sensitivity of the Plus network. We also found that the primary tilt is well-constrained at both sensitivities given favourable spin configuration (in-plane spins, high precession).

Motivated by the prospects of using spins to infer the formation history of BBH mergers, we analyzed two injection grids in which we varied the spin tilts while fixing the masses and spin magnitudes to values reflecting two classes of hierarchical mergers: 1G+2G and 2G+2G. In the 1G+2G grid, we found that the Plus network can provide much more meaningful spin constraints for the 2G primary component, especially with regards to the spin magnitude. The Plus network constrained $a_1\gtrsim 0.5$ at 95\% credibility for around half of the injections. With Design, we could only obtain similar constraints for 16\% of the injections. The primary tilt was consistently well-measured at both sensitivities for aligned and misaligned orientations. The combination of a well-constrained magnitude and tilt gained with the Plus network will allow us to make more confident statements about how individual binaries formed. Furthermore, we applied a cost function to gauge the precision and accuracy of our spin measurements, finding that the Plus network can yield constraints with on average 40\% lower cost than Design across the full set of injected 1G+2G signals. The largest improvements come from binaries with primary spin tilted at or slightly beyond $90^\circ$.

The results for the 2G+2G grid show that the spins of such systems are considerably more difficult to accurately recover, even at Plus sensitivity. Although constraints on the secondary spin improve relative to the 1G+2G grid, the primary spin is less well-measured, as the two component spins become degenerate with each other for near-equal masses. We conclude that the 1G+2G binaries are better candidates for extracting stronger spin measurements using the Plus network relative to Design, and we expect these results to be generalizable to other types of mergers between BHs of different generations.

Additional work is needed to understand what conditions are needed to measure both spins accurately. Most likely, this will require more sensitive detectors. Though we only considered near-future detector upgrades that will come online during this decade, third-generation gravitational-wave detectors, including LIGO Voyager \citep{LIGO:2020xsf}, Cosmic Explorer \citep{Reitze:2019iox}, and Einstein Telescope \citep{Maggiore:2019uih}, will provide even greater signal fidelity. We leave the topic of spin resolution using these next-generation detectors to future work, complementing \citet{Vitale:2016icu} and similar studies.

Finally, we analyzed a simulated population sampled from astrophysical BBH population models. The correlation between high mass ratio and spin recovery with the Plus network is robust to more realistic variations in the masses and spins, however we do not find greater improvement with the Plus network among high-spinning binaries like we did for the mass-spin grid. We find that the Plus network consistently provides better constraints on the primary spin for $q>2$ sources, which is consistent our mass-spin results. Although spin uncertainties are overall larger compared to the injection grids, we still find that the Plus network allows us to confidently identify spin misalignment at $1.5-2$ higher rates than Design for injections with substantial in-plane spin.

We considered a relatively simple population in which the mass and spin distributions were independent. However, because the masses and spins are both influenced by formation channels, a more detailed population might include multiple subpopulations of sources with different mass and spin properties, and would serve as another interesting extension of our study. 

\begin{acknowledgements}

We thank Scott Oser for suggesting the use of a cost function to measure the accuracy of parameter recovery. We also thank the LIGO parameter estimation working group for helpful discussions, including Patricia Schmidt for providing comments on this manuscript.

This material is based upon work supported by NSF’s LIGO Laboratory (\url{www.ligo.caltech.edu}) which is a major facility fully funded by the National Science Foundation (\url{www.nsf.gov}). We gratefully acknowledge the support of both the LIGO Laboratory and Compute Canada (\url{www.computecanada.ca}) for provision of computing resources. This work was also supported by the Natural Sciences and Engineering Research Council of Canada (\url{www.nserc-crsng.gc.ca}) through the Discovery Grants and CGS Master's programs.

\end{acknowledgements}

\appendix

\section{Cost function}\label{app:cost}

\begin{figure}[t!]
\epsscale{1.15}
\plotone{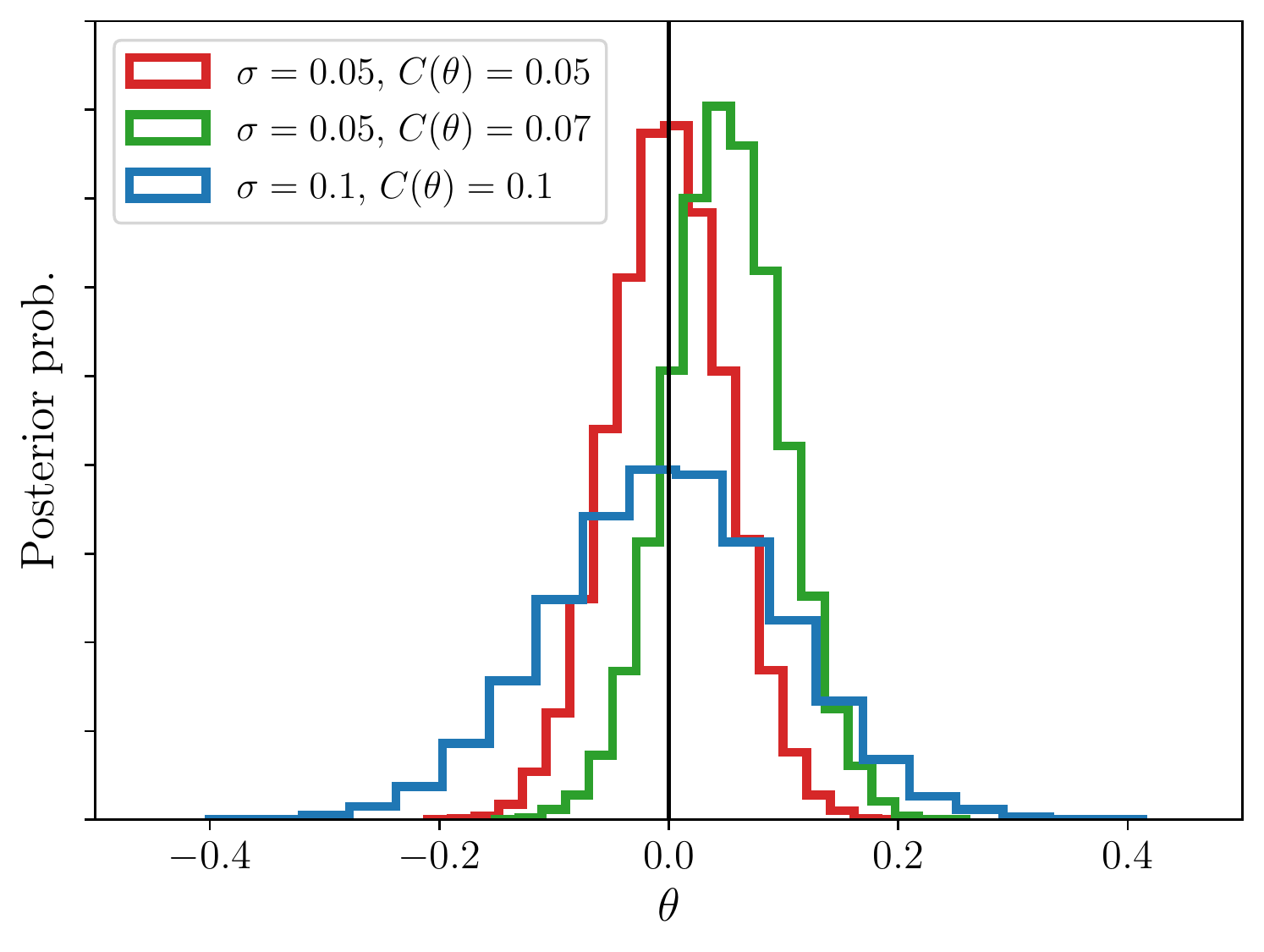}
\caption{Illustrative synthesized posteriors and their cost functions. Each distribution is sampled from a Gaussian with some mean, $\mu$, and standard deviation, $\sigma$. The red distribution has $\sigma=0.05$ and $\mu=0$, green has $\sigma=0.05$ and $\mu=0.05$, and blue has $\sigma=0.1$ and $\mu=0$. The cost functions are computed using $\theta_0=0$. \label{fig:cost_example}}
\end{figure}

To provide further intuition regarding the cost function introduced in Section \ref{sec:cost}, we show illustrative examples of synthesized posterior distributions with different cost functions in Figure \ref{fig:cost_example}. The posteriors are sampled from Gaussian distributions with certain means and standard deviations. The cost function is computed with respect to $\theta_0=0$ (taken to be the injected value in this example). In the case where the posterior mean is the same as the injected value, the cost function simply returns the standard deviation of the posterior. If the posterior is offset from the injected value, as is the case for the green distribution with mean $\theta=0.05$ (so an offset of $0.05$), the cost function is larger than the standard deviation. In fact, the cost function is equivalent to the root sum of the variance and the offset squared, $C(\theta)=\sqrt{\sigma^2+{\rm offset}^2}\approx 0.07$. 

In terms of our ability to infer the source properties of binary coalescences, the cost function presents different information about our inferences compared to the widths of the posteriors/CIs. Even if the population were drawn from a distribution that exactly matched the prior, we would naturally expect some fraction of the events (for a given credible level) to not have their parameters accurately recovered due to the occurrence of statistically improbable noise realizations. For our analyses, we assigned each injected signal a numeric seed that was used to generate the noise at both Design and Plus sensitivity. This was done to ensure that the noise realizations had the same probabilities across the two networks. Thus, if a posterior is offset from the injected value at Design sensitivity, the cost function will be able to capture whether that offset improves, worsens, or stays the same as a result of the enhanced sensitivity provided by the Plus network, and weigh that alongside any changes in the precision of the measurement (i.e.~the width of the posterior). This way, the cost function can be a useful tool for flagging cases where the spin parameters are not more {\it accurately} recovered with the Plus network compared to Design, even when the {\it precision} of the measurement increases (see Figure \ref{fig:2G2G_corner} and the surrounding discussion).

\begin{figure}[t!]
\epsscale{1.15}
\plotone{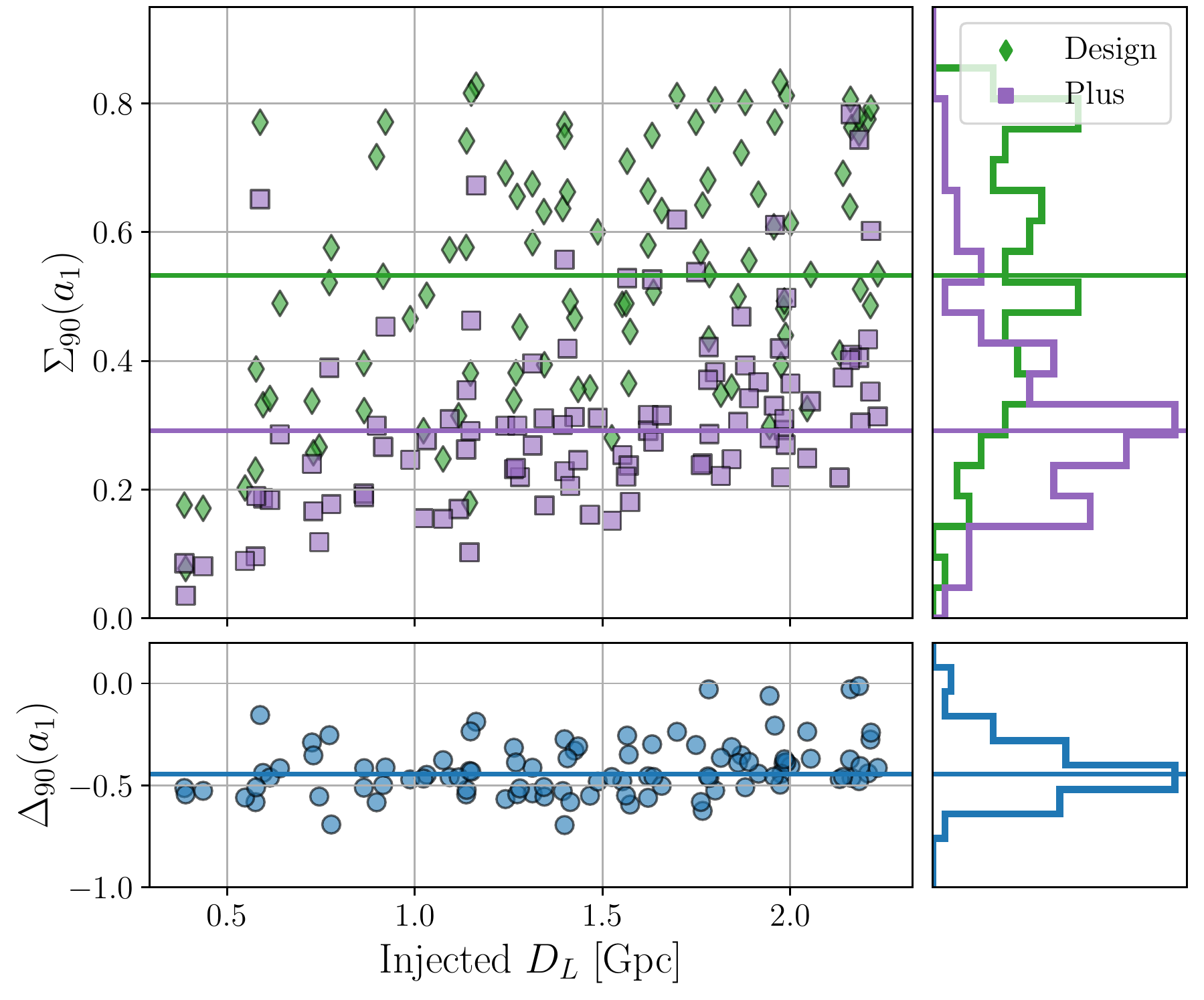}
\caption{Distribution of constraints on the primary spin magnitude, $a_1$, as a function of luminosity distance, $D_L$, for the 1G+2G spin tilt grid. {\it Top:} Widths of the 90\% CIs, $\Sigma_{90}(a_1)$. The horizontal lines mark the medians of $\Sigma_{90}(a_1)$. {\it Bottom:} Relative change in CI widths between Design and Plus sensitivity, $\Delta_{90}(a_1)$, also with a histogram. \label{fig:1G2G_a_1_sigma}}
\end{figure}

\begin{figure}[t!]
\epsscale{1.15}
\plotone{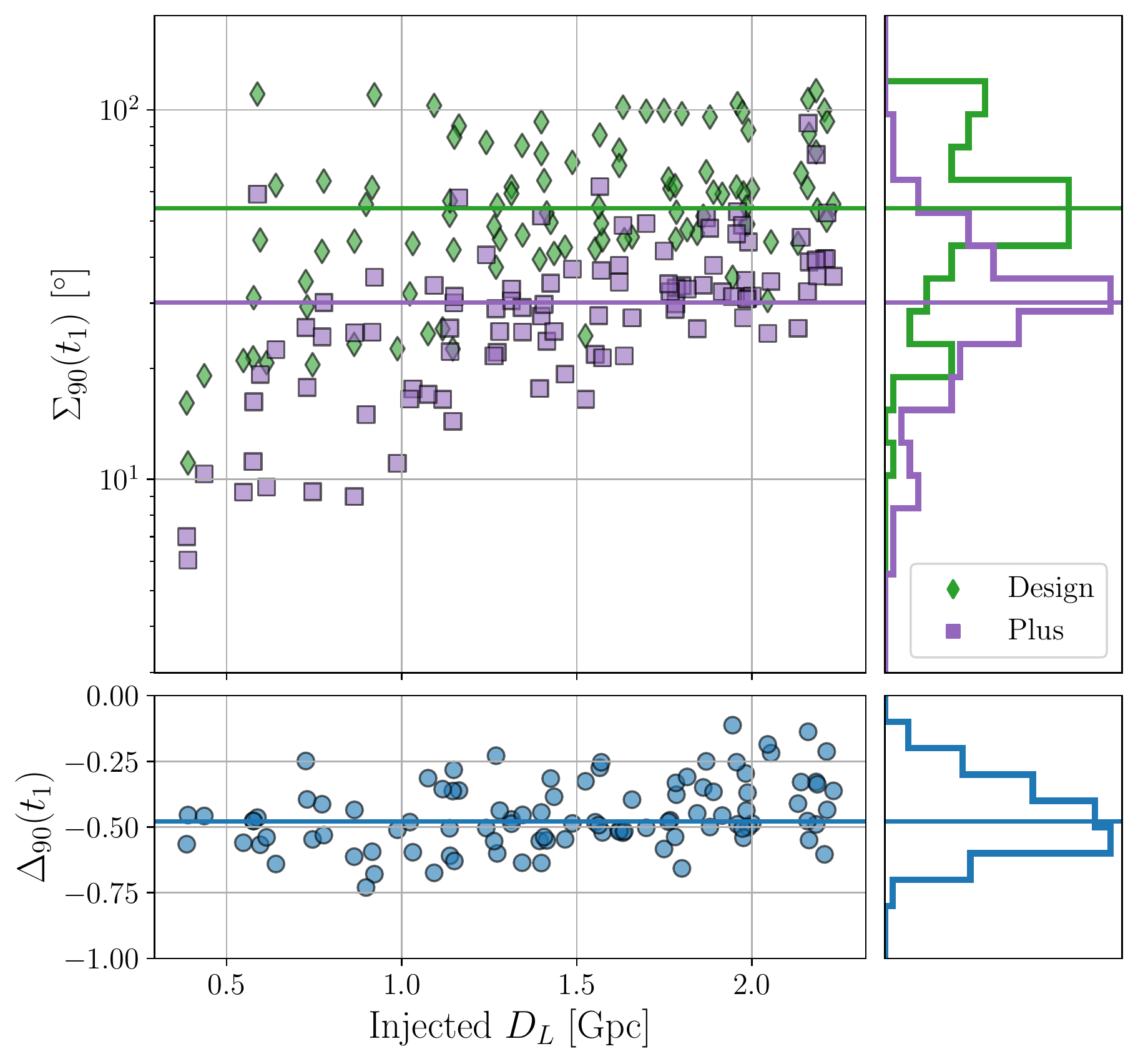}
\caption{Distribution of constraints on the primary spin tilt, $t_1$, as a function of luminosity distance, $D_L$, for the 1G+2G spin tilt grid. {\it Top:} Widths of the 90\% CIs, $\Sigma_{90}(t_1)$. The horizontal lines mark the medians of $\Sigma_{90}(t_1)$. {\it Bottom:} Relative change in CI widths between Design and Plus sensitivity, $\Delta_{90}(t_1)$, also with a histogram. \label{fig:1G2G_t_1_sigma}}
\end{figure}

\begin{figure}[t!]
\epsscale{1.15}
\plotone{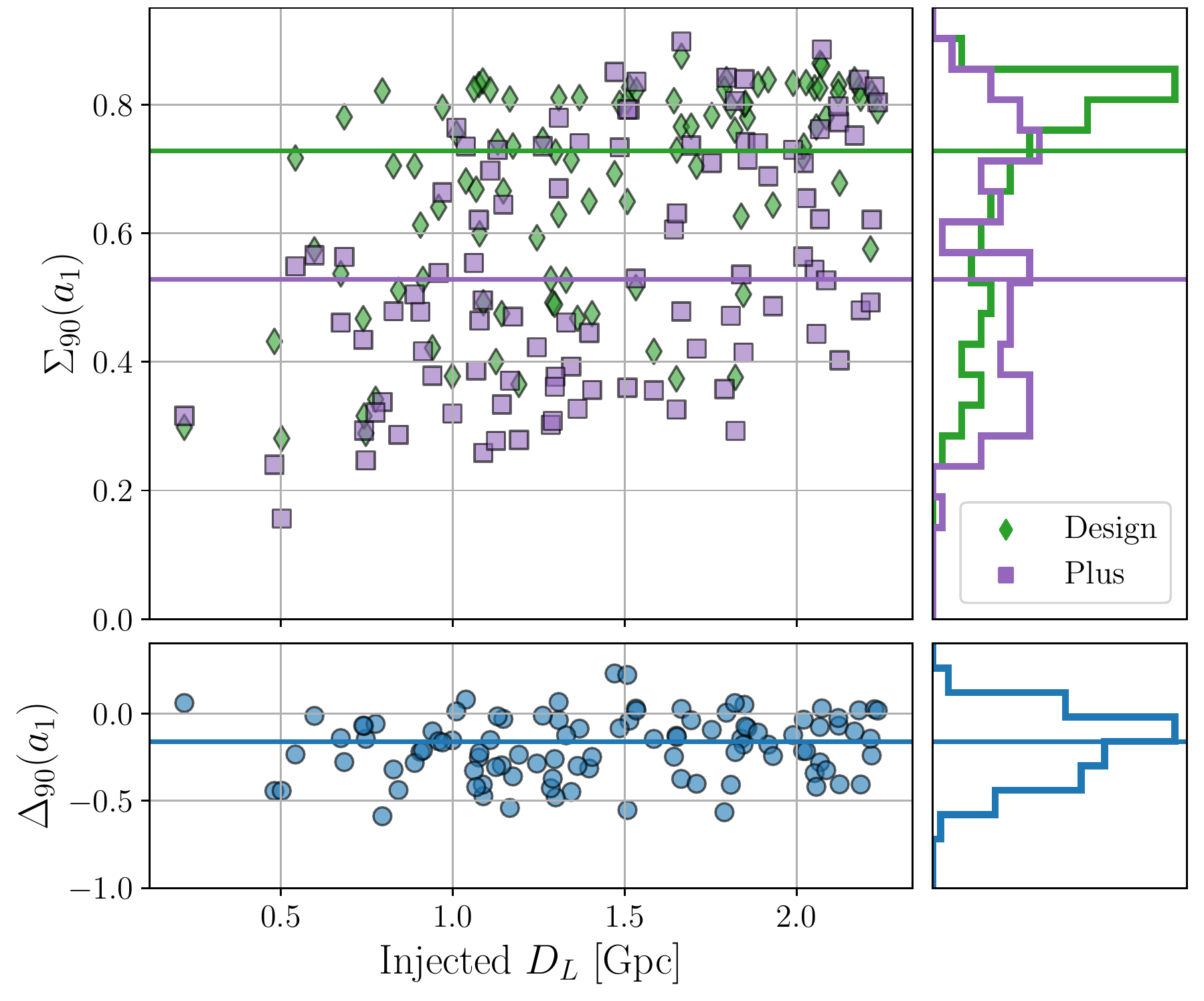}
\caption{Distribution of constraints on the primary spin magnitude, $a_1$, as a function of luminosity distance, $D_L$, for the 2G+2G spin tilt grid. {\it Top:} Widths of the 90\% CIs, $\Sigma_{90}(a_1)$. The horizontal lines mark the medians of $\Sigma_{90}(a_1)$. {\it Bottom:} Relative change in CI widths between Design and Plus sensitivity, $\Delta_{90}(a_1)$, also with a histogram. \label{fig:2G2G_a_1_sigma}}
\end{figure}

\begin{figure}[t!]
\epsscale{1.15}
\plotone{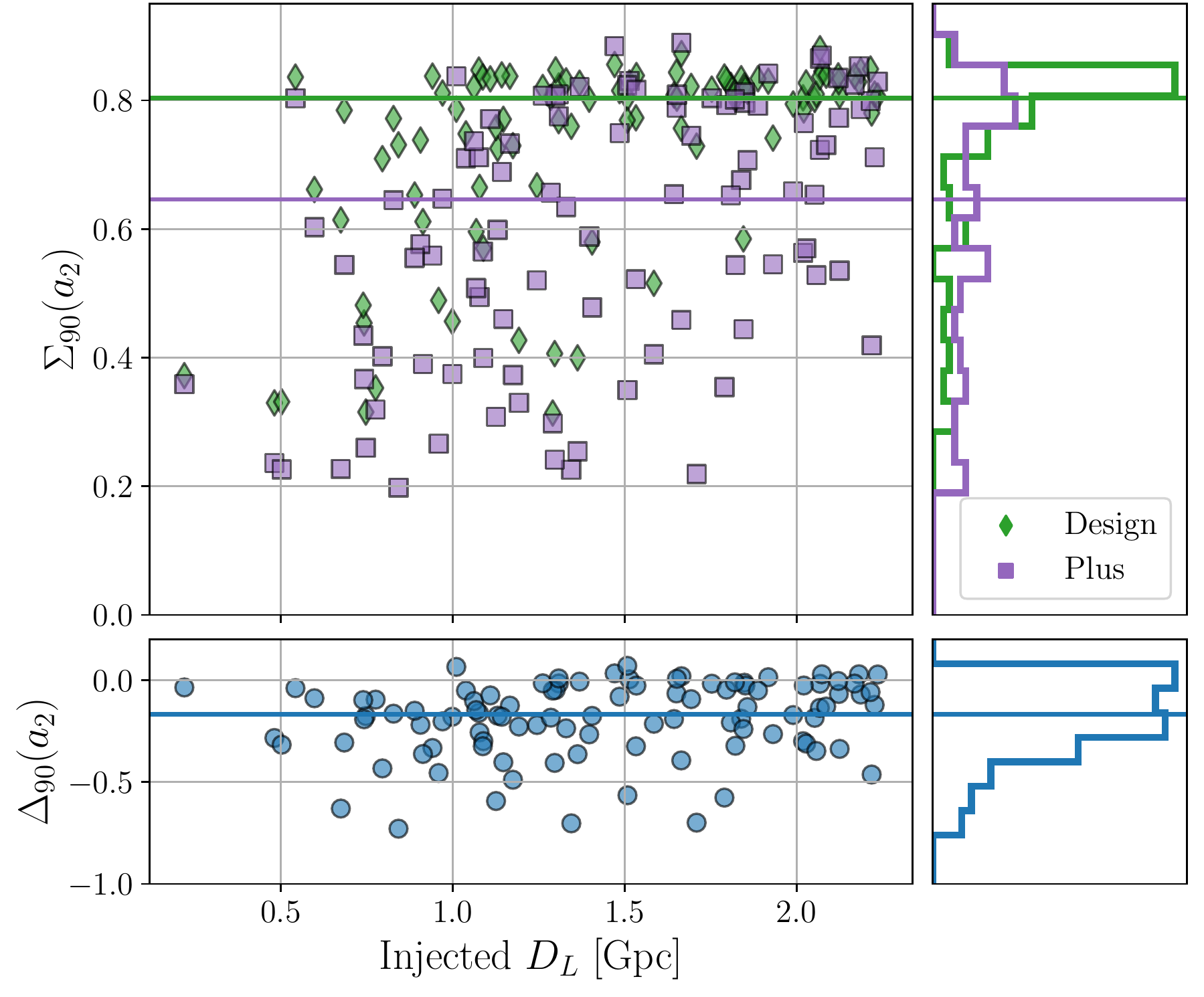}
\caption{Same as Figure \ref{fig:2G2G_a_1_sigma}, showing constraints on the secondary spin magnitude, $a_2$. \label{fig:2G2G_a_2_sigma}}
\end{figure}

\begin{figure}[t!]
\epsscale{1.15}
\plotone{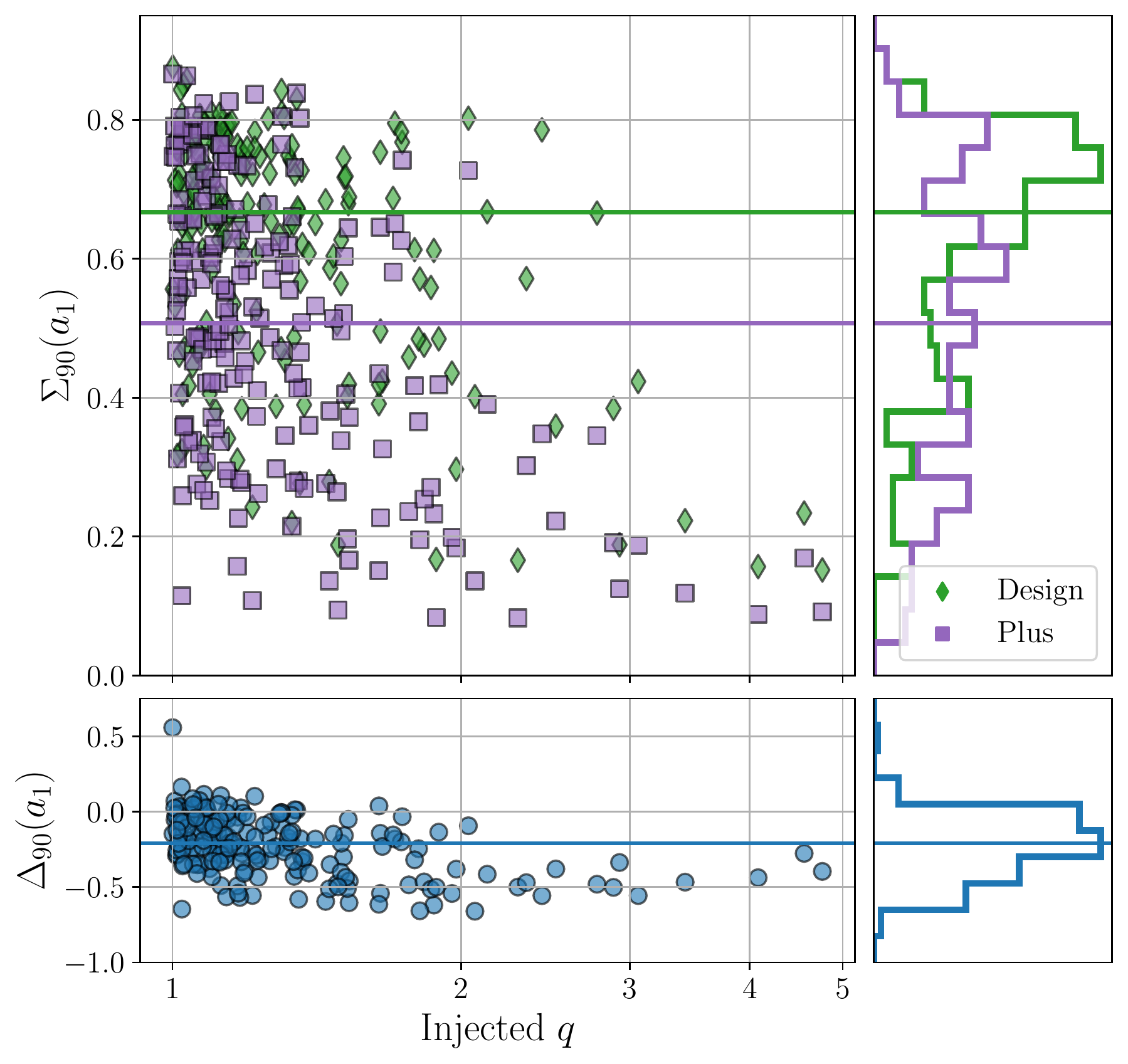}
\caption{Distribution of constraints on the primary spin magnitude, $a_1$, as a function of mass ratio, $q$, for the population injection set. {\it Top:} Widths of the 90\% CIs, $\Sigma_{90}(a_1)$. The horizontal lines mark the medians of $\Sigma_{90}(a_1)$. {\it Bottom:} Relative change in CI widths between Design and Plus sensitivity, $\Delta_{90}(a_1)$, also with a histogram. \label{fig:pop_a_1_sigma}}
\end{figure}

\begin{figure}[t!]
\epsscale{1.15}
\plotone{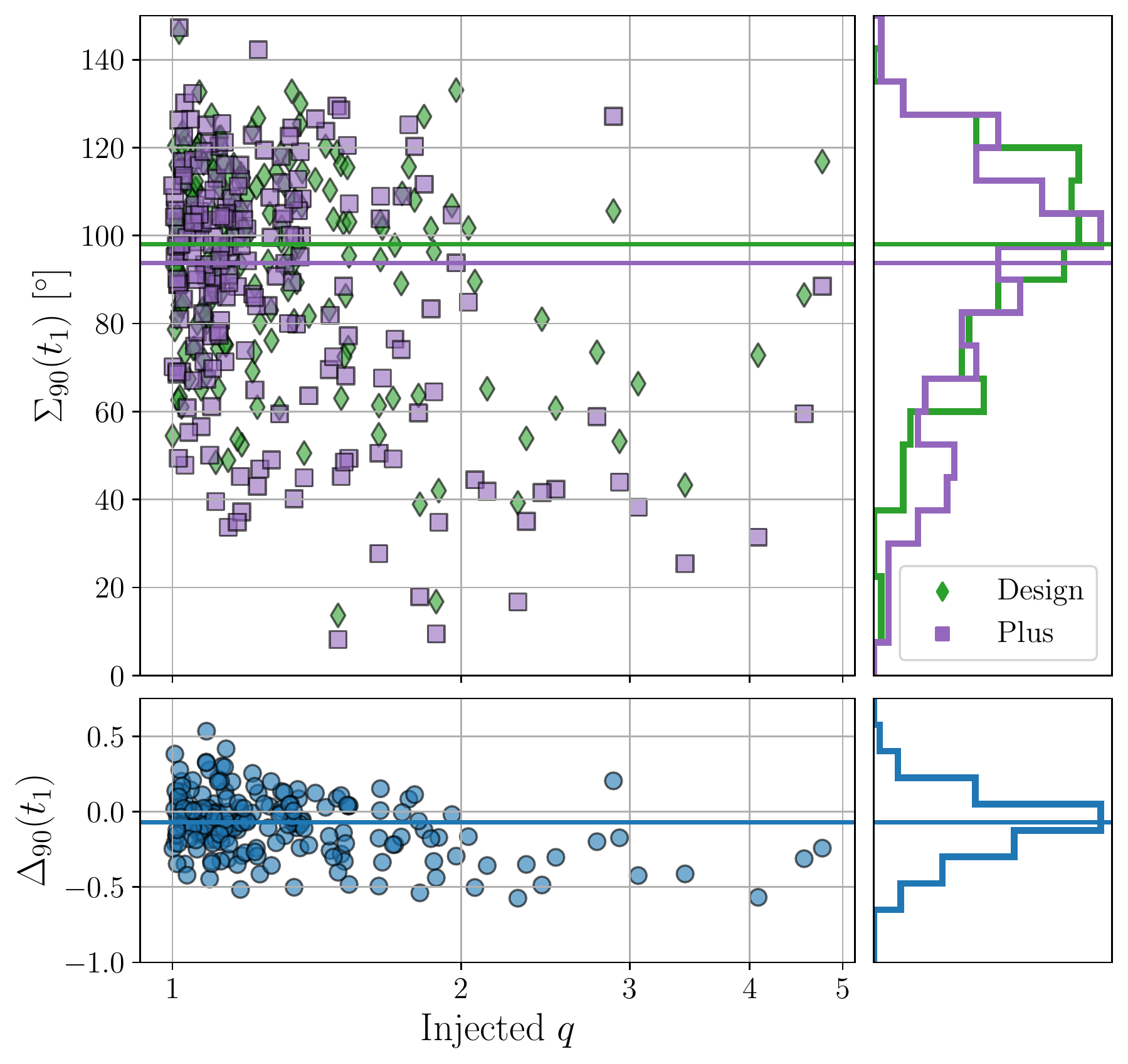}
\caption{Distribution of constraints on the primary spin tilt, $t_1$, as a function of mass ratio, $q$, for the population injection set. {\it Top:} Widths of the 90\% CIs, $\Sigma_{90}(t_1)$. The horizontal lines mark the medians of $\Sigma_{90}(t_1)$. {\it Bottom:} Relative change in CI widths between Design and Plus sensitivity, $\Delta_{90}(t_1)$, also with a histogram. \label{fig:pop_t_1_sigma}}
\end{figure}

\section{Posterior widths}\label{app:widths}

The cost function combines together information about the position of the posterior peak relative to the true injected value and its width. In cases where the posterior is significantly offset from the injected value of a parameter, the cost function will be much larger than the standard deviation. Regardless, the posterior width alone, representing the precision of the measurement, may still be of interest. Here, we compile a set a of figures showing the widths of the 90\% CIs and their relative change between Design and Plus sensitivity, $\Sigma_{90}(\theta)$ and $\Delta_{90}(\theta)$, on the vertical axis instead of $C(\theta)$ and $\Delta C(\theta)$. When discussing results in terms of the CI widths, our overall assessment of the injection studies does not significantly change, though we maintain the cost function still provides useful information for the reasons described above.

Beginning with the spin tilt 1G+2G grid, in Figures \ref{fig:1G2G_a_1_sigma}--\ref{fig:1G2G_t_1_sigma} we show the widths of the 90\% CIs for the primary spin. At Plus sensitivity, 50\% of the injections have $a_1$ constrained at $\Sigma_{90}<0.3$. 50\% of the injections also have $t_1$ constrained at  $\Sigma_{90}<30^\circ$. Similar to the cost function, the widths of the primary spin CIs decreases by nearly 50\% on average between Design and Plus sensitivity. 

Figures \ref{fig:2G2G_a_1_sigma}--\ref{fig:2G2G_a_2_sigma} show the widths of the 90\% CIs for the primary spin for the 2G+2G spin tilt grid. Looking at the medians again, 50\% of the injections have $a_1$ constrained at $\Sigma_{90}<0.35$, whereas for $a_2$, 50\% have it constrained at $\Sigma_{90}<0.65$. Our previous conclusion that the spins of 1G+2G binaries are better constrained than in 2G+2G binaries, which was based on the cost function, still holds when looking at the widths of the constraints themselves. Despite the slightly better $a_1$ constraints than for $a_2$, we do observe cases where Plus sensitivity produces a worse constraint (in the sense of a wider posterior) than Design, whereas all the Plus network $a_2$ constraints are narrower than or equal to the Design constraint. For instance, this occurs for the nearest signal located at around 200 Mpc, but from the cost function in Figure \ref{fig:2G2G_a1} we can see that despite the slightly worse precision, the peak is closer to the injected value at Plus sensitivity than Design, resulting in a lower cost. We also note that, like the cost function, the spin constraints for 2G+2G binaries has a weaker correlation with distance compared to 1G+2G binaries.

Primary spin constraints for the population injections are shown in Figures \ref{fig:pop_a_1_sigma}-\ref{fig:pop_t_1_sigma}. When analyzing the constraints using the cost function, we found modest improvements for the $a_1$ constraints, and minimal improvements for the $t_1$ constraints, which remains true when looking at the posteriors widths. Similarly, systems with unequal masses have their constraints narrow by a greater amount between Design and Plus sensitivity.

\bibliography{main}{}
\bibliographystyle{aasjournal}



\end{document}